\documentclass[12pt,oneside,reqno]{amsart}

\usepackage{geometry}
\usepackage{amsmath}
\usepackage{enumerate}
\usepackage{enumitem}
\usepackage{kantlipsum}
\allowdisplaybreaks
\counterwithout{footnote}{section}
\usepackage{cleveref}
\DeclareMathOperator{\Tr}{Tr}
\usepackage{setspace}
\usepackage{graphicx}

\usepackage{cleveref}
\usepackage{multirow}
\usepackage[normalem]{ulem}
\usepackage{amsmath,amssymb,amsfonts}
\usepackage{mathtools,leftindex,tensor,mhchem}
\usepackage{amsthm}
\newtheorem*{remark}{Remark}
\usepackage{enumerate}
\usepackage{pifont}
\usepackage{mathrsfs}
\usepackage{bm}
\usepackage{xcolor}
\usepackage{textcomp}
\usepackage{accents}
\usepackage{manyfoot}
\usepackage{algorithm}
\usepackage{algorithmicx}
\usepackage{algpseudocode}
\usepackage{listings}
\newtheorem{definition}{Definition}[section]

\newtheorem{prop}[definition]{Proposition}

\newtheorem{assump}[definition]{Assumption}
\newtheorem{consis}[definition]{Consistency Check}
\newtheorem{invaria}[definition]{Gauge Invariance Check}
\usepackage{comment}

\def\eq#1{Eq.~$(\ref{#1})$}

\def\S{{\mathscr{S}}}

\def\({\textup{(}}
\def\){\textup{)}}

\begin{document}

\title[Euclidean Axioms]{Axioms for Quantum Yang-Mills Theories - 1. Euclidean Axioms}

\author[1]{Min Chul Lee}
\email{min.lee@chch.ox.ac.uk}

\begin{abstract}
    This paper extends the notion of Schwinger functions to quantum Yang-Mills theories and proposes the axioms they should satisfy. Two main features of this axiom scheme are that we assume existence of gauge-invariant co-located Schwinger functions and impose physical properties only on them. This is in accordance with the fundamental principle of gauge theories that only gauge-invariant quantities can be physical observables.
\end{abstract}

\maketitle

\tableofcontents

\section{Introduction}
\label{sec:intro}

In this paper we adjust the existing Euclidean axiom scheme \cite{OstSch73, OstSch75, Nel73, time79} for quantum field theory to axiomatize most non-Abelian gauge theories (a.k.a Yang-Mills theories) in current usage. A typical construction program starts with the Euclidean framework and derives the Minkowski theory as a consequence. Therefore, if one seeks to axiomatize gauge theories, proposal of Euclidean axioms is the reasonable starting point.

We discuss two most widely recognized strategies for construction of quantum gauge theories:
\begin{itemize}

\item
The first approach is the lattice approximation. The continuum infinite volume limit was established for the 2D Abelian Higgs model in a series of papers \cite{BryFroSei79, BryFroSei80, BryFroSei81}. The mass generation of the gauge field (Higgs mechanism) is also proven. Furthermore, existing axioms of \cite{OstSch73, OstSch75}, except for cluster decomposition, are rigorously verified. The Abelian Higgs model has been rigorously constructed in 3D finite volume by T. Balaban, whose renormalization scheme was later improved by J, Dimock; refer to \cite{Dimoscalar} and references therein. Dimock also has been working on 3D spinor QED, where he established ultraviolet stability and regularity through a series of papers \cite{Dim18, Dim20, Dim20a}. Regarding lattice approximation of Yang-Mills theories, Balaban wrote a large series of papers, as described in the expository articles \cite{Dimo1, Dimo2, Dimo3} and references therein such as \cite{Bal11,Bal12, Bal13, Bal14}. Balaban's main achievement was the introduction of length scale dependent block averaging. In fact, Dimock took up this set of ideas in analyzing 3D spinor QED.

\item
The second approach is stochastic quantization. In \cite{GroKinSen89}, a functional measure in the axial gauge is constructed as a solution of the 2D Yang-Mills stochastic differential equation. \cite{stoc1} addresses the SPDE for 2D Yang-Mills theory as well, with inclusion of the Higgs field. However, rather than working in a fixed gauge, \cite{stoc1} rigorously constructs the gauge orbit space of physical states and analyzes its topology and establishes a Markov process invariant under the gauge action. Moreover, it conjectures existence of a unique gauge-invariant functional measure associated with this Markov process. In 3D, the Yang-Mills-Higgs model is considered in the paper \cite{stoc2}. Again, the paper rigorously defines the gauge orbit space and obtains a Markov process as the solution of Yang-Mills SPDE. Still, existence of a unique gauge-invariant functional measure is not proven yet.

\item There also exist papers such as \cite{Dri89} in which both of the above approaches are addressed. It shows that lattice approximation and stochastic quantization yield the same result in 2D under the complete axial gauge.

\end{itemize}

What is clear from the above discussion is that there may be multiple paths of constructing gauge theories. Our axioms are focused on the non-Abelian cases and formulated to be independent of which approach is chosen, and designed to apply to the final construction only.  

 In fact, there exists general consensus within the constructive quantum field theory community on what properties should be required for a sensible gauge theory~\cite{jaffecommun}, most of which have been explicitly stated and verified on lattice setting~\cite{OstSei78}. This paper may be regarded as systematic statement of such requirements in continuum $\mathbb{R}^4$ with fully constructed theories in consideration. This is in accordance with history of development of constructive QFT. The Wightman and Haag-Kastler axioms preceded construction of non-trivial examples in the history of rigorous quantum field theory, as described in the article \cite{Jaf00}. The axioms were suggested in early 1960s but first rigorous construction of nontrivial examples was achieved in late 1960s through the $\phi^4$ and Yukawa theories in $1+1$ spacetime dimensions; more difficult examples such as the $\phi^4$ theory in $3$ Euclidean dimensions were constructed during 1970s and 1980s. Therefore, we hope that this paper will serve as a guide post for mathematical construction of quantum Yang-Mills theories.

\section{Notations}

We closely follow the notations~\cite[Section 2 and 6]{OstSch73} and~\cite[Chapter 15]{Wei96}. 

\begin{definition}
    Throughout this paper, the expression 
\[
C:=D
\]
means that $C$ is defined as $D$. The terms  ``Yang-Mills theory" and  ``Non-Abelian gauge theory" are used interchangeably.
\end{definition}

\begin{definition}
  \label{coordinate basic}
   The symbol $\undertilde{x}$ denotes a point in $\mathbb{R}^4$ whose coordinates with respect to the standard basis are given by $(x^0,x^1,x^2,x^3)$. Following~\cite[p.87]{OstSch73}, $\undertilde{\theta x}$ is defined as a point in $\mathbb{R}^4$ with the coordinates $(-x^0,x^1,x^2,x^3)$.

   We also use notations $\undertilde{y}$, $\undertilde{w}$ and $\undertilde{z}$ for the same purpose. The indices $i,j \in \{0,1,2,3\}$ label vector components in $\mathbb{R}^4$ with respect to the standard basis.\footnote{As we are in the Euclidean space, it may be more natural for vector indices to take values in $\{1,2,3,4\}$. However, we follow the notations of~\cite{OstSch73, time79} to maintain consistency with existing literature.} Moreover, we assume the Euclidean metric on $\mathbb{R}^4$ with respect to the standard basis throughout this paper.

\end{definition}

\begin{definition}
For any $N \in \mathbb{N}$, $\S(\mathbb{R}^{4N})$ denotes the space of complex-valued Schwartz functions on $\mathbb{R}^{4N}$. $\S(\mathbb{R}^{4N})$ is a nuclear Fr\'echet space, with an example of the semi-norms explicitly presented in~\cite[p.86]{OstSch73}. Denote by $\S'(\mathbb{R}^{4N})$ the space of temperate distributions equipped with the strong dual topology. 

We use the notation $\mathscr{S}_{\mathbb{R}}(\mathbb{R}^{4N})$ for the space of real-valued Schwartz functions.

\end{definition}

\begin{definition}
\label{lie group and lie algebra}
     The gauge group $G$ denotes a finite-dimensional simple compact (real) Lie group, typically $SU(N)$ for a positive integer $N \geq 2$.\footnote{In this paper, we only consider gauge particles which are their own antiparticles, such as gluons in QCD. That is, the gauge fields $A_{\alpha i}$ appearing in Section~\ref{motive heuristics} are \textit{Hermitian} according to the physics terminology. The restriction of $G$ to a simple compact Lie group is another simplification. Generalization to arbitrary products of simple compact gauge groups (and $U(1)$) as well as non-Hermitian gauge fields (such as W$^{\pm}$ bosons) are, at least conceptually, straightforward. However, there may be some technical subtleties with $U(1)$, as described in footnote~\ref{xxx} below.} We fix a bi-invariant Riemannian metric on $G$ throughout this paper.\footnote{From Schur's lemma, such a metric on $G$ is unique up to multiplication by a positive real number.} The Lie algebra associated with $G$ is denoted as $\mathfrak{g}$ and we write its Lie bracket as $[ \, , \, ]$.  
     
The Peter-Weyl Theorem states that such $G$ may be realized as a matrix Lie group and the Lie bracket $[ \, , \, ]$ as the commutator of matrices. Such representations are assumed throughout this paper via specification of the matter and gauge fields in a given theory. Following~\cite[Chapter 15]{Wei96}, we fix a set of generators for $\mathfrak{g}$ 
     \[
     \bigl \{ t_\alpha \mid \alpha = 1,2, \cdots, \text{dim } \mathfrak{g} \bigr \}
     \]
     which are normalized in the sense that $\Tr(t_\alpha t_\beta)= \delta_{\alpha \beta}$.
See Definition~\ref{gauge index def} for more details.

Furthermore, denote as $[ \, , \, ]_{\otimes}$ the bilinear mapping from $\mathfrak{g} \times \mathfrak{g}$ into $\mathfrak{g} \otimes \mathfrak{g}$ defined by
\[
[ t_{\alpha} , t_\beta ]_{\otimes} := t_\alpha \otimes t_\beta - t_\beta \otimes t_\alpha
\]
   \end{definition}

\begin{definition}
\label{evaluation def}
  Let $\mathcal{M}$ be the space of smooth real-valued bounded functions $\mathbb{R}^4$ with all partial derivatives rapidly decaying. It is easy to see as in~\cite{484755} that such $\mathcal{M}$ is the direct sum of $\mathscr{S}_{\mathbb{R}}(\mathbb{R}^4)$ and $\mathbb{R}$. It is therefore a nuclear Fr\'echet space. For any $k \in \mathbb{N}$, denote by $\mathcal{M}^{\otimes k}$ the $k$-fold tensor product of $\mathcal{M}$ with the topology as in~\cite[Chapter 43]{Treves67}.\footnote{Since $\mathcal{M}$ is a nuclear Fr\'echet space, there is only one reasonable topology on $\mathcal{M}^{\otimes k}$ according to~\cite[Theorem 50.1]{Treves67}.} 

Assume that $\mathfrak{g}$ is a real matrix Lie algebra. For any $k, k' \in \mathbb{N}$, $\mathcal{M}^{\otimes k} \otimes g^{\otimes k'}$ may be identified with the space of $g^{\otimes k}$-valued mappings with the entries (with respect to any chosen basis of $g^{\otimes k}$) being elements of $\mathcal{M}^{\otimes k}$. Therefore, $\mathcal{M}^{\otimes k}_{\mathfrak{g}}$ is just a finite direct sum of $\mathcal{M}^{\otimes k}$. The continuous linear mapping $\mathscr{P}_{k,k'} : \mathcal{M}^{\otimes k} \otimes \mathfrak{g}^{\otimes k'} \to \mathcal{M}^{\otimes k} \otimes \mathfrak{g}$ is defined to be the unique linear extension of the following formula:
  \begin{equation}
      \label{evaluation C}
         \mathscr{P}_{k,k'}\Bigl( F_1 \otimes \cdots \otimes F_{k'} \Bigr) :=  F_1 \cdots F_{k'}
  \end{equation}
   for $F_1 \cdots, F_{k'} \in \mathcal{M}^{\otimes k} \otimes \mathfrak{g}$, where the expression $F_1 \cdots F_{k'}$ should be understood as composition of linear maps (or product of matrices). From now on, subscripts are omitted from $_{k,k'}$ as they will be clear from the context. Note that 
   \[
   \mathscr{P} \bigl( [t_\alpha, t_\beta]_\otimes \bigr)=[t_\alpha, t_\beta].
   \]

We denote the composition $\Tr \circ \mathscr{P}$ as $\Tr_\otimes$. Note that
 \begin{equation}
    \label{noncolocated trace}
        \Tr_\otimes \bigl( F_1 \otimes \cdots \otimes F_{k'}  \bigr)=  \Tr\bigl(F_1 \cdots  F_{k'}\bigr) \in \mathcal{M}^{\otimes k}
    \end{equation}
    where $F_1 \cdots, F_{k'} \in \mathcal{M}^{\otimes k} \otimes \mathfrak{g}$. In particular,     \[
      \Tr_{\otimes}( t_\alpha \otimes t_\beta )=\Tr_{\otimes}( t_\beta \otimes t_\alpha )=\delta_{\alpha \beta}.
    \]

    Everything stated so far in this definition may be extended to complex-valued functions and complexification of $\mathfrak{g}$, and the same notations will be used for both real and complex cases. In fact, we assume from now on that $\mathfrak{g}$ is complexfied. This is in order to address the Schwinger functions in the next section as elements of $\mathscr{S}'(\mathbb{R}^{4N})$ for both the gauge and matter fields.

   \end{definition}

   \begin{definition}
   \label{local gauge mapping def}
    Let $g  : \mathbb{R}^4 \to G $ be a smooth mapping. Its differential $Dg$ is than a mapping of $\undertilde{x} \in \mathbb{R}^4$ such that
    \[
    D_{\undertilde{x}} g \in L\bigl(\mathbb{R}^4, T_{g(\undertilde{x})} G \bigr)
    \]
     Using the Euclidean metric on $\mathbb{R}^4$ and the bi-invariant Riemannian metric on $T_{g(\undertilde{x})} G$ assumed in Definition~\ref{lie group and lie algebra}, we may define the operator norm $\lVert \cdot \rVert_{op}$ on $L\bigl(\mathbb{R}^4, T_{g(\undertilde{x})} G \bigr)$. Note that $\lVert \cdot \rVert_{op}$ does not depend on $\undertilde{x}$ by construction. We say that $\undertilde{x} \to D_{\undertilde{x}}g$ is rapidly decaying at infinity if
\[
\sup_{\undertilde{x} \in \mathbb{R}^4} \bigl(1 + \lVert \undertilde{x} \rVert \bigr)^n \lVert D_x g \rVert_{op} < \infty
\]
for all $n \in \mathbb{N}$. It is straightforward to generalize this notion of rapid decay at infinity to higher-order differentials of $g$.  

In order to address the local action of $G$ in the context of temperate distribution, we introduce the following group:
\[
\mathcal{G}:=\Bigl\{ g : \mathbb{R}^4 \to G \, \bigl \lvert \, g\text{ is smooth with the differential of each order rapidly decaying at infinity} \Bigr\}
\]   
where the group operation is defined point-wise. That is, $(g_1 g_2)(\undertilde{x}):=g_1(\undertilde{x}) g_2(\undertilde{x})$ and $g^{-1}(\undertilde{x}):=[g(\undertilde{x})]^{-1}$. 

Since $G$ is a compact Lie group, $\mathcal{G}$ is indeed a group under such operations. Moreover, it is not difficult to observe that entries of $g \in \mathcal{G}$ with respect to any representation\footnote{More specifically, any choice of basis for the representation space} of $G$ are elements of $\mathcal{M}$.\footnote{Such $\mathcal{G}$ may be too restrictive, but it is the best we can find for the purpose of defining co-located products of Schwinger functions corresponding to the gauge fields. We do not exclude the possibility of a more singular local gauge structure than $\mathcal{G}$ and any modifications required accordingly. In fact, such adjustments are needed to address the free $U(1)$ theory in the Lorentz gauge, which is already a well-established result. See Section~\ref{u(1) example}. \label{xxx}}

\end{definition}

\begin{definition}
\label{gauge subtlety}
Throughout this paper, $A : (\mathbb{R}^{4})^2 \to \mathbb{C}$ denotes a generic element of complex-valued $\mathcal{M}^{\otimes 2}$. $A \bigl \lvert_{\text{diag}} : \mathbb{R}^4 \to \mathbb{C}$ is the function $\undertilde{x} \to A(\undertilde{x},\undertilde{x})$.

Such $A$ will play a crucial role in consistency of the gauge structure for co-located Schwinger functions.
\end{definition}

\begin{definition}
\label{gauge index def}
    We assume that the matter fields are partitioned into multiplets to furnish representations of the gauge group $G$ in a given theory. More specifically, let $\mathcal{R}$ be the labeling of all multiplets in the theory. Then, there exists a representation of $G$ for each $r \in \mathcal{R}$ such that the $r$-multiplet of matter fields are components with respect to a given (ordered) basis in the representation space. We denote the representation space by $V_r$ and the basis by $\{e_{k_r}\}$; here the index $k_r$ takes the value from $\{1, 2, \cdots, \text{dim } V_r\}$.
    
    We assume in addition that the adjoint\footnote{Here the term \textit{adjoint} refers to the dual spinor representation. For example, it denotes (Euclidean) Dirac adjoint in the case of a (Euclidean) Dirac spinor.} of the matter fields in the $r$-multiplet forms a multiplet in the representation of $G$ dual to that of $r$ with respect to the dual basis of $\{e_{k_r}\}$. Such multiplet is labeled by $\overline{r}$. For any $g \in \mathcal{G}$, the notation $g_{(r)}$ is used to make it explicit the representation under which the values of $g$ are expressed.  
    
    Note that, by definition, $V_{\overline{r}}$ is the dual space of $V_r$ and $\{e_{k_{\overline{r}}}\}$ is the dual basis for $\{e_{k_r}\}$. We assume further that $\{e_{k_{\overline{r}}}\}$ is ordered in the same way as $\{e_{k_r}\}$ so that $\langle e_{k_{\overline{r}}}, e_{k_r} \rangle_{V_{\overline{r}} \times V_r}=\delta_{k_r k_{\overline{r}}}$ and the adjoint of the matter field component corresponding to $e_{k_r}$ is the component in the $\overline{r}$-multiplet corresponding to $e_{k_{\overline{r}}}$. We further assume the following completeness relation:
    \[
   \sum_{k_r, k'_{\overline{r}}} \delta_{k_r k'_{\overline{r}}}\bigl\langle \phi, e_{k_r} \bigr \rangle_{V_{\overline{r}} \times V_r } \bigl \langle e_{k'_{\overline{r}}} , v \bigr \rangle_{V_{\overline{r}} \times V_r } = \bigl \langle \phi , v \bigr \rangle_{V_{\overline{r}} \times V_r } \text{ for all } \phi \in V_{\overline{r}} \text{ and } v \in V_r
    \]

    For notational convenience, we may also use the symbols $V^*_r$ and $\{e^*_{k_r}\}$ to denote the dual space $V_{\overline{r}}$ and the dual basis $\{e_{k_{\overline{r}}}\}$ respectively. Note that $e^*_{k_r}=\sum_{k'_{\overline{r}}}\delta_{k_r k'_{\overline{r}}}e_{k'_{\overline{r}}}$.

\end{definition}

\begin{definition}
\label{spinor index def}
    For the spinor indices of matter fields, we modify~\cite[Section 6]{OstSch73}. All fields in each $r$-multiplet are assumed to be of the same spinor character. As such, the index $\nu_r$ describes the (Euclidean) spinor character of the fields in the $r$-multiplet. By construction, the index $\nu_{\overline{r}}$ corresponds to the representation of $SO(4)$ dual to that of $\nu_r$.
\end{definition}

\begin{definition}
    $\Psi_{k_r \nu_r}$ denotes the matter field which is the component of the basis element $e_{k_r}$ in the $r$-multiplet, so that the whole multiplet may be expressed as $\Psi_{v_{r}}=\sum_{k_r}\Psi_{k_r \nu_r}e_{k_r}$. With the notations from Definition~\ref{gauge index def} and Definition~\ref{spinor index def}, the adjoint of the field $\Psi_{k_r \nu_r}$ is $\Psi_{k_{\overline{r}} \nu_{\overline{r}} }$ and vice versa.\footnote{As a concrete example, let us consider (the Euclidean version of) QCD~\cite[Section 18.7]{Wei96}, where quark fields and their Dirac adjoints are the matter fields. In this case, $\mathfrak{R}=\{ u, \overline{u}, c,\overline{c}, t, \overline{t}, d,\overline{d}, s,\overline{s},b,\overline{b} \}$ corresponds to the flavors, where $u$ denotes the $u$-quark \textit{field} while $\overline{u}$ is its Dirac adjoint, and similarly for other flavors. 
    With $r$ denoting any element of $\{ u,c,t,d,s,b\}$, $k_r$ takes three values (= colors), furnishing the fundamental representation $\bm{3}$ of $SU(3)$. $k_{\overline{r}}$ takes three values as well, furnishing the dual representation $\overline{\bm{3}}$.} 
\end{definition}

\begin{definition}
\label{euclidean rep}
   For $U,V \in SU(2)$, we denote by $S(U,V)$ the representation of $SO(4)$ corresponding to a given spinor index and $R=R(U,V)$ for the fundamental representation of $SO(4)$. This is identical to the notations in~\cite[p.102]{OstSch73}.
\end{definition}

\begin{definition}
\label{index and permutation}
For any $m \in \mathbb{N} \cup \{0\}$\footnote{For brevity, we suppress $m$ in index sets unless all elements of a given set is explicitly written out. Nevertheless, cardinality of index sets will always be clear from the context.}
\begin{itemize}
    \item The index set $I$ denote any ordered collection of $m$ tuples of the forms $(k_r, v_r)$ and $(\alpha,i)$.\footnote{That is, $I$ labels the fundamental fields (both matter and gauge) in a given theory.}

    \item The index set $\mathcal{I}$ denotes any ordered collection of $m$ tuples of the forms $(\nu_r, \nu'_{\overline{r}})$ and $(\alpha,i,j)$.\footnote{As presented in subsequent sections, $\mathcal{I}$ will label the gauge-invariant composite matter fields and the field strength tensor derived from gauge potentials.}

\item The index set $\mathfrak{I}$ denotes any ordered collection of $m$ tuples of the forms $(\nu_r, \nu'_{\overline{r}})$ and $(i,j,i',j')$.\footnote{As presented in subsequent sections, $\mathfrak{I}$ will label the composite fields that are fully gauge-invariant.}
\end{itemize}
let $I$ denote any ordered set of $n$ pairs of $(k_r, v_r)$ or $(\alpha,i)$
For example, $I$ may take the following forms when $m=3$:
\[
 \bigl\{  (\alpha_1, i_1), (\alpha_2, i_2) ,(\alpha_3, i_3) \bigr\} \text{ and any permutation of the pairs}
\]
   \[
    \bigl\{ (k_{r}, \nu_{r}), (\alpha_1, i_1), (\alpha_2, i_2) \bigr\} \text{ and any permutation of the pairs}
   \] 
\[
\bigl\{ (k_{r_1}, \nu_{r_1}),  (k_{r_2}, \nu_{r_2}), (\alpha, i) \bigr\} \text{ and any permutation of the pairs}
\]
  \[
  \bigl\{ (k_{r_1}, \nu_{r_1}),  (k_{r_2}, \nu_{r_2}), (k_{r_3}, \nu_{r_3}) \bigr\} \text{ and any permutation of the pairs}.
  \]
For any permutation $\sigma$ of $n$ elements, let $\sigma \cdot I$, $\sigma \cdot \mathcal{I}$ and $\sigma \cdot \mathfrak{I}$ be the ordered index set obtained by permuting according to $\sigma$ the elements of $I$, $\mathcal{I}$ and $\mathfrak{I}$ respectively.

From now on, summation convention will be assumed between any pair of repeated indices.
\end{definition}

\begin{definition}
\label{schwartz sequence def}
Let $\uline{f}=\bigl( \uline{f}_{m,I} \bigr)$ be a sequence enumerated by $m \in \mathbb{N} \cup \{0\}$ and all possible $I$ as in Definition~\ref{index and permutation} such that \footnote{This is adjustment of the notations in~\cite[pp.97-98]{time79} to encompass spacetime and gauge indices, by referring to~\cite[p.103]{OstSch73}.} 
\begin{itemize}
       \item $ \uline{f}_{m, I} \in \mathscr{S}(\mathbb{R}^{4m})$ with $\uline{f}_{0, \emptyset} \in \mathbb{C}$
    \item all but finitely many elements are zero
    \item For $m \neq 0$, the support of $\uline{f}_{m, I}$ is contained in $\bigl\{ (\undertilde{x}_1, \cdots \undertilde{x}_{m} ) \mid x^0_1, \cdots ,x^0_{m} \geq 0 \bigr\}$
\end{itemize}
Let $\Theta$ be an involution mapping defined component-wise on $\uline{f}$ as follows:
\begin{equation}
\begin{split}
    \label{involution theta}
\uline{f} \to \Theta \uline{f} \quad \text{ by } \quad \Theta \uline{f} := \bigl( [\Theta \uline{f}]_{m,I} \bigr) \text{ with }  [\Theta \uline{f}]_{m,I}(\undertilde{x}_1, \cdots \undertilde{x}_{m} ):=\overline{\uline{f}_{m,I^*}(\undertilde{\theta x}_{m}, \cdots \undertilde{\theta x}_{1} )}
    \end{split}
\end{equation}
where $\undertilde{\theta x}$ is defined in Definition~\ref{coordinate basic} and $I^*$ is defined in the same way as~\cite[Formula (6.2)]{OstSch73} but with gauge indices included. For example,
\[
I^*= \bigl\{ (\alpha, i),  (k_{\overline{r_2}}, \nu_{\overline{r_2}}), (k_{\overline{r_1}}, \nu_{\overline{r_1}}) \bigr\} \text{ for }I=\bigl\{   (k_{r_1}, \nu_{r_1}), (k_{r_2}, \nu_{r_2}),(\alpha, i) \bigr\}.
\]

We may make similar definitions with $I$ replaced by $\mathcal{I}$ or $\mathfrak{I}$.  Lastly, the reflection operator $\Theta$ with respect to the zeroth coordinate is defined in the same ways as in~\cite[p.87]{OstSch73}.
\end{definition}

\begin{definition}
    Let $V$ be a finite-dimensional complex vector space. Then, $\mathscr{S}(\mathbb{R}^{4N}) \otimes V$ is the space of $V$-valued Schwartz functions on $\mathbb{R}^{4N}$. 
    For any $f \in \mathscr{S}(\mathbb{R}^{4N}) \otimes V$ and $T \in \mathscr{S}'(\mathbb{R}^{4N})$, we use the notation $T(f)$ to denote an element of $V$ obtained by $T$ acting on the $\mathscr{S}(\mathbb{R}^{4N})$ part of $f$.

    More concretely, let us fix a basis $\{ e_l\}$ of $V$ and write $f=f_l e_l$, where $f_l \in \mathscr{S}(\mathbb{R}^{4N})$. Then, $T(f):=T(f_l)e_l \in V$.
\end{definition}

\begin{definition}
\label{regularizing scheme}
Denote by $\{ \Delta_n \}_{n=1}^\infty$ a sequence of elements in $\mathscr{S}(\mathbb{R}^{4 \times 3})$ that converges to the restriction onto the thin diagonal in the weak$^*$ limit of temperate distributions. That is, for any $F \in \mathscr{S}(\mathbb{R}^{4 \times 3})$, we have
\[
\lim\limits_{n \to \infty} \int_{(\mathbb{R}^4)^3} \Delta_n(\undertilde{x},\undertilde{y},\undertilde{z}) F(\undertilde{x},\undertilde{y},\undertilde{z})dxdydz = \int_{\mathbb{R}^4} F(\undertilde{x},\undertilde{x},\undertilde{x})dx.
\]

For $f \in \mathscr{S}(\mathbb{R}^4)$, we use the notation $\Delta_n(f)$ to denote an element of $\mathscr{S}(\mathbb{R}^{4 \times 2})$ defined by
\begin{equation}
    \label{deltaf}
    \Delta_n(f):= \int_{\mathbb{R}^4} \Delta_n(\cdot,\cdot,\undertilde{z})f(\undertilde{z})dz.
\end{equation}

We may add the gauge representation indices for matter fields. More specifically, let $ \bigl\{ \Delta_{n, k_r k'_{\overline{r}}} \bigr\} \subset \mathscr{S}(\mathbb{R}^{4 \times 3})$ be such that
\[
\lim\limits_{n \to \infty} \int_{(\mathbb{R}^4)^3} \Delta_{n, k_r k'_{\overline{r}}}(\undertilde{x},\undertilde{y},\undertilde{z}) F(\undertilde{x},\undertilde{y},\undertilde{z})dxdydz = \delta_{ k_r k'_{\overline{r}}}\int_{\mathbb{R}^4} F(\undertilde{x},\undertilde{x},\undertilde{x})dx.
\]
for any $F \in \mathscr{S}(\mathbb{R}^{4 \times 3})$. $\Delta_{n, k_r k'_{\overline{r}}}(f)$ is defined in the same way as~\eq{deltaf}.
\end{definition}

\section{Motive Heuristics and Illustrations}
\label{motive heuristics}

With the notations presented in the previous section, we first provide a heuristic outline of the main ideas, which will be established with full mathematical rigor in subsequent sections and the next paper.

In~\cite{OstSch73, OstSch75, time79}, Schwinger functions are vacuum expectation values of field operators for the given theory.\footnote{Note that \cite{OstSch73, OstSch75} define Schwinger functions only at non-coinciding arguments while \cite{time79} include coinciding arguments as well. Refer to \cite[p.371]{coinciding} for a more detailed comparison. In this paper, we follow the approach of \cite{time79} since composite operators at coinciding arguments must be addressed for gauge invariance.} However, a crucial feature of gauge theories is the presence of the local action by a gauge group $G$, resulting in gauge redundancy. That is, not all state vectors have physical meaning.

Hence, the notion of vacuum expectation values must be generalized. More specifically, suppose that we are given the following:
\begin{itemize}
    \item A vector space $H$ of all possible states equipped with a conjugate-linear form $\langle \, , \rangle$ and containing a state vector $\Omega$

    \item A collection of matter fields $\Psi_{ k_r \nu_r}$ and gauge fields $A_{\alpha i}$\footnote{These field operators must be \textit{interacting}. It is well-known that (perturbative) renormalizability of a gauge theory depends on gauge fixing. Here we just assume that renormalized (or interacting) field operators under a certain gauge fixing are given, and present the axioms they should satisfy. At least, all field operators in an equivalence class under the gauge action yield the same gauge-invariant co-located products as in~\eq{gauge invariant schwinger} by construction. A caveat here is that local gauge symmetry may not be transitive at the level of interacting field operators like $\Psi_{ k_r \nu_r}$ and $A_{\alpha i}$. In that case, there may exist multiple different spaces of the state vectors and collections of gauge-invariant co-located products corresponding to each gauge equivalence class (= orbit). We conjecture that this non-uniqueness issue is somehow related to existence of multiple superselection sectors~\cite[p.108]{Haa96} as well as spontaneous symmetry breaking~\cite[Chapter 19]{Wei96}. Actual construction of field operators and detailed analysis of such properties are the major outstanding issues for the future.} which are temperate distributions whose values are operators acting on $H$
\end{itemize}

Note that $H$ is not necessarily a (pre-)Hilbert space since $\langle \, , \rangle$ is not assumed to be positive (semi-)definite.\footnote{~\cite[Ch.10]{BoLoOkTo90} assumes that $H$ is a Hilbert space and $\langle \, , \rangle$ is expressed in terms of the inner product of $H$. However, such assumptions seem too restrictive and we work in a more general setting, which is still in the same spirit as~\cite[Ch.10]{BoLoOkTo90} since it also focuses mostly on the indefinite metric and the physical Hilbert space will eventually be obtained via taking quotient. Later, we analytically continue from Euclidean to Minkowski spacetime and establish an extended version of the Reconstruction Theorem~\cite[Theorem 3.7]{StrWig64} encompassing a local gauge structure, where all relevant details will be presented with full rigor. For now, we proceed more heuristically, as stated out in the start of this section.} However, $\Omega$ is later interpreted as a physical vacuum when restricted to a subspace of $H$, which will be called a physical Hilbert space of observable states.

Even without structure of a Hilbert space, we may still define a Schwinger function by the formula\footnote{Elitzur's theorem~\cite{Eli75, FroMorStr81} states that the expectation value of gauge-noninvariant field operators with respect to a gauge-invariant functional measure vanishes identically. This result is originally stated for lattice but clearly holds in continuum limit as well, provided that the limit exists. Such a restriction is bypassed by introducing a gauge-fixing functional into the path integral, as originally proposed by Faddeev and Popov~\cite{FadPop67} in the heuristic level on continuum and rigorously verified on lattice in~\cite{FroMorStr81, OstSei78}. As shown in Section~\ref{squant}, the conjugate-linear form $\langle \, , \rangle$ together with the state space $H$ correspond to path integral for a given gauge theory via the moment problem. Therefore, we are led to the conclusion that they must be constructed under a specific gauge fixing condition in order to avoid Elitzur's theorem.}
\begin{equation}
\begin{split}
\label{schwinger vev}
    &S_{I}\bigl(\undertilde{x}_1, \cdots,  \undertilde{x}_{m+l} \bigr) := \Bigl \langle \Omega, \Psi_{ k_{r_1} \nu_{r_1}}(\undertilde{x}_1) \cdots \Psi_{ k_{r_m} \nu_{r_m}}(\undertilde{x}_m) A_{\alpha_1i_1}(\undertilde{x}_{m+1}) \cdots  A_{\alpha_li_l}(\undertilde{x}_{m+l}) \Omega \Bigr \rangle \\
    &\text{with }I \text{ given here as }\bigl\{( k_{r_1} \nu_{r_1}), \cdots, ( k_{r_m} \nu_{r_m}),(\alpha_1,i_1), \cdots, (\alpha_l,i_l)\bigr\} 
\end{split}
\end{equation}
or any permutation of the matter fields and gauge fields. By the condition that field operators $\Psi_{ k_r \nu_r}$ and $A_{\alpha i}$ are operator-valued temperate distributions, such Schwinger functions are elements of $\mathscr{S}'\bigl(\mathbb{R}^{4(m+l)}\bigr)$. They are required to satisfy the original axioms in~\cite{OstSch73, OstSch75} \textit{except for} physical properties such as reflection positivity and cluster decomposition.

Let us now consider the local gauge action more in detail. Following~\cite[Formulas (1.61) and (1.72)]{Fra08} with the (physical) coupling constant absorbed into field operators, such an action of $G$ on the field operators is given by the formulas:
\begin{equation}
\label{local action def}
\begin{cases}
     \bigl(g \cdot \Psi \bigr)_{k_r \nu_r}(f):=\Psi_{k'_r \nu_r}\Bigl( f(\cdot)\bigl \langle e^*_{k_r} \circ  g_{(r)}(\cdot), e_{k'_r} \bigr \rangle_{V^*_r \times V_r} \Bigr)\\   t_{\alpha} \bigl[ g \cdot A \bigr ]_{\alpha j}(f):=  A_{ \alpha j}\Bigl(fgt_\alpha g^{-1} \Bigr) - \int_{\mathbb{R}^4} \bigl[(\partial_j g) g^{-1} \bigr](\undertilde{x}) f(\undertilde{x}) dx
\end{cases}
\end{equation}
where $f \in \mathscr{S}(\mathbb{R}^4)$ and $g \in \mathcal{G}$. In the second formula of~\eq{local action def}, $g$ and $t_\alpha$ are assumed to be in the adjoint representation.

~\eq{schwinger vev} naturally leads to the local action of $G$ on Schwinger functions as follows:
\begin{equation}
\begin{split}
\label{schwinger gauge1}
&\bigl[g \cdot S \bigr]_{I} \bigl(\undertilde{x}_1, \cdots, \undertilde{x}_{m+l} \bigr)\\
&:= \Bigl \langle \Omega, \bigl [g \cdot \Psi \bigr ]_{k_{r_1} \nu_{r_1}}(\undertilde{x}_1) \cdots \bigl [g \cdot \Psi \bigr ]_{k_{r_m} \nu_{r_m}}(\undertilde{x}_m) \bigl[ g \cdot A \bigr ]_{\alpha_1,i_1}(\undertilde{x}_{m+1}) \cdots   \bigl[g \cdot A \bigr]_{\alpha_l,i_l}(\undertilde{x}_{m+l}) \Omega \Bigr \rangle
\end{split}
\end{equation}
or similarly for any permutation of indices. Using~\eq{local action def}, we may express~\eq{schwinger gauge1} without explicit resort to field operators, which will be be the formal definition of local gauge action for Schwinger functions in the next section. Here, we give simple examples for the matter fields and gauge fields respectively:
\begin{itemize}
    \item In the presence of a single matter field and its adjoint only,
    \begin{equation}
    \label{matter field local action example}
   \bigl[g \cdot S \bigr]_{(k_{r},\nu_{r}), (k'_{\overline{r}},\nu'_{\overline{r}})}(f \otimes h):=S_{(k''_{r},\nu_{r}) ,(k'''_{\overline{r}},\nu'_{\overline{r}})}\Bigl( f(\cdot)\bigl \langle e^*_{k''_r} \circ  g_{(r)}(\cdot), e_{k_r} \bigr \rangle_{V^*_r \times V_r} \otimes h(\cdot)\bigl \langle e^*_{k'''_{\overline{r}}} \circ  g_{(\overline{r})}(\cdot), e_{k'_{\overline{r}}} \bigr \rangle_{V_{\overline{r}}^* \times V_{\overline{r}} }\Bigr)
\end{equation}
where $f,h \in \mathscr{S}(\mathbb{R}^4)$ and the natural identifications $V^*_{\overline{r}}=V_r$ and $e^*_{k_{\overline{r}}}=\delta_{k_{\overline{r}} k'_r}e_{k'_r}$ are assumed.

\item In the presence of two gauge fields only,
\begin{equation}
\begin{split}
    \label{gauge field local action example}
    &\bigl(t_\alpha \otimes t_{\alpha'}\bigr)\bigl[g \cdot S \bigr]_{(\alpha, i), (\alpha', i')}(f \otimes h)\\
    &:=S_{(\alpha, i), (\alpha', i')}\Bigl( f gt_{\alpha}g^{-1} \otimes h gt_{\alpha'} g^{-1} \Bigr)-S_{(\alpha, i)}\Bigl(f gt_\alpha g^{-1}\Bigr)\otimes \Biggl(\int_{\mathbb{R}^4} h (\partial_{i'} g)g^{-1} \Biggr) \\
    &-\Biggl(\int_{\mathbb{R}^4} f (\partial_{i} g)g^{-1}\Biggr) \otimes S_{(\alpha', i')}\Bigl(h gt_{\alpha'} g^{-1}\Bigr)+\Biggl(\int_{\mathbb{R}^4} f (\partial_{i} g)g^{-1}\Biggr)\otimes\Biggl(\int_{\mathbb{R}^4} h (\partial_{i'} g)g^{-1}\Biggr)
    \end{split}
    \end{equation}
where $f,h \in \mathscr{S}(\mathbb{R}^4)$.
\end{itemize}

The fundamental principle of gauge theories is that all physical observables must be gauge-invariant. According to~\cite{OstSch73, OstSch75}, reflection positivity is a key properties of Schwinger functions that lead to positivity of Wightman functions, which in turn makes it possible to reconstruct the underlying Hilbert space, cf. \cite[Thereom 3.7]{StrWig64}. Therefore, reflection positivity may be regarded as a physical requirement and must be imposed on gauge-invariant quantities only. Similarly, cluster decomposition is related to locality and therefore only gauge-invariant observables need to satisfy this property.

For this purpose, we should construct gauge-invariant Schwinger functions starting from~\eq{schwinger vev}. Among local field operators, the simplest gauge-invariant ones\footnote{We retain locality for Schwinger functions. However, non-local quantities such as the Wilson loop will be derived from such Schwinger functions in a later section.} for the matter fields is of the following form:

\begin{equation}
\label{matter field composite}
\bigl \lvert \Psi \bigr \rvert^2_{\nu_r  \nu_{\overline{r}}'} (\undertilde{x} ) := \delta_{k_r k_{\overline{r} }'   } \bigl(\Psi_{k_r \nu_r} \Psi_{k_{\overline{r}} '\nu_{\overline{r}}'}\bigr)(\undertilde{x})
\end{equation}
provided that the co-located product is well-defined.

For the gauge fields $A_{\alpha i}$, situation is more involved. Following~\cite[Formula (1.78)]{Fra08} or~\cite[Formula (15.1.13)]{Wei96}, the field strength tensor is defined as:
\begin{equation}
\label{gauge field composite0}
   t_\alpha F_{\alpha i j}(\undertilde{x}) := t_\alpha \Bigl(\partial_i A_{\alpha j}(\undertilde{x}) - \partial_j A_{\alpha i}\Bigr)(\undertilde{x})+ [t_\beta,t_\gamma]\Bigl(A_{\beta i} A_{\gamma j} \Bigr)(\undertilde{x})
\end{equation}
which leads to the gauge-invariant local operator
\begin{equation}
\label{gauge field composite}
 F^2_{iji'j'}(\undertilde{x}):= \Bigl(F_{\alpha i j}  F_{\alpha i' j'}\Bigr)(\undertilde{x})=\Tr\Bigl( F_{\alpha ij}t_{\alpha} \otimes F_{\alpha' i' j'}t_{\alpha'}\Bigr)(\undertilde{x})
\end{equation}
where $\Tr$ is formally used here for co-located field operators. Of course, we must assume that such co-located products in~\eq{gauge field composite0} and~\eq{gauge field composite} are somehow well-defined.

Using~\eq{matter field composite} and~\eq{gauge field composite0}, we may introduce the following intermediate form of Schwinger functions:
\begin{equation}
\begin{split}
\label{intermediate schwinger}
&\mathcal{S}_{\mathcal{I} }\bigl(\undertilde{x}_1,  \cdots, \undertilde{x}_{m+l} \bigr) :=\Bigl \langle \Omega, \bigl \lvert \Psi \bigr \rvert^2_{\nu_{r_1}  \nu_{\overline{r_1}}'}(\undertilde{x}_1 ) \cdots \bigl \lvert \Psi \bigr \rvert^2_{\nu_{r_m}  \nu_{\overline{r_m}}'}(\undertilde{x}_m )  F_{\alpha_1i_1j_1}(\undertilde{x}_{m+1}) \cdots  F_{\alpha_l i_lj_l}(\undertilde{x}_{m+l}) \Omega \Bigr \rangle\\
& \text{with }\mathcal{I} \text{ given here as } \bigl\{ (\nu_{r_1}, \nu'_{\overline{r_1}}), \cdots,(\nu_{r_m}, \nu'_{\overline{r_m}}), (\alpha_1, i_1,j_1), \cdots, (\alpha_l, i_l,j_l) \bigr\}
\end{split}
\end{equation}
or again any permutation of the field operators and corresponding indices.~\eq{intermediate schwinger} is the non-Abelian version of the formula appearing in~\cite[p.383 Corollary 4.6]{BryFroSei81}.\footnote{The sum in the formula of Schwinger functions appearing the cited paper must be replaced by a product, which is a typo confirmed via email correspondence with the authors.} 

However,~\eq{intermediate schwinger} is still not fully gauge-invariant due to the non-Abelian structure. Rather, one may heuristically compute that
\begin{equation}
\begin{split}
\label{intermediate schwinger gauge action}
&\bigotimes_{k=1}^l t_{\alpha_k}\bigl[g \cdot \mathcal{S}_{\mathcal{I} }\bigr]\bigl(f_1 \otimes  \cdots \otimes f_{m+l} \bigr)  \\
&=\Bigl \langle \Omega, \bigl \lvert \Psi \bigr \rvert^2_{\nu_{r_1}  \nu_{\overline{r_1}}'}(f_1 ) \cdots \bigl \lvert \Psi \bigr \rvert^2_{\nu_{r_m}  \nu_{\overline{r_m}}'}(f_m )  F_{\alpha_1i_1j_1}\bigl( f_{m+1} gt_{\alpha_1}g^{-1} \bigr) \cdots  F_{\alpha_l i_lj_l}\bigl( f_{m+l} gt_{\alpha_l}g^{-1} \bigr) \Omega \Bigr \rangle
\end{split}
\end{equation}
where $f_1, \cdots, f_{m+l} \in \mathscr{S}(\mathbb{R}^4)$ and the tensor product of matrices with operator-valued entries, where the usual multiplication of scalars is replaced by composition of functions for the entries, is assumed for $F$'s. In fact, \eq{intermediate schwinger gauge action} is equivalent to the transformation rule
\[
t_\alpha [g \cdot F]_{\alpha i j} = g t_\alpha g^{-1} F_{\alpha i j}
\]
for the field strength tensor $F$, which is well-established at the classical level. Therefore, with the transformation rule for a two-fold tensor product of $F_{\alpha i j}$ is given by
\begin{equation}
\label{F2 rule}
\bigl(t_\alpha \otimes t_{\alpha'}\bigr) \bigl[g \cdot F \bigr]_{\alpha i j}(\undertilde{x})\bigl[g \cdot F \bigr]_{\alpha' i' j'}(\undertilde{x'}) = \Bigl(F_{\alpha i j}(\undertilde{x})g (\undertilde{x})t_\alpha  g^{-1}(\undertilde{x}) \Bigr)\otimes \Bigl(F_{\alpha' i' j'}(\undertilde{x'})g (\undertilde{x'})t_\alpha  g^{-1}(\undertilde{x'}) \Bigr).
\end{equation}
Then, just like~\eq{gauge field composite}, we again formally apply the mapping $\Tr$ to~\eq{F2 rule} to compute that
\begin{equation}
\begin{split}
\label{F2 gauge invariance heuristic}
\Tr\Bigl( \bigl[g \cdot F\bigr]_{\alpha ij}t_{\alpha} \otimes \bigl[ g \cdot F \bigr]_{\alpha' i' j'}t_{\alpha'}\Bigr)(\undertilde{x}) 
&= \bigl(F_{\alpha i j} F_{\alpha' i' j'}\bigr)(\undertilde{x}) \Tr\Bigl( g(\undertilde{x})t_{\alpha}g^{-1}(\undertilde{x})g(\undertilde{x})t_{\alpha'}g^{-1}(\undertilde{x})\Bigr)\\
&=\Tr\Bigl( F_{\alpha ij}t_{\alpha} \otimes F_{\alpha' i' j'}t_{\alpha'}\Bigr)(\undertilde{x})
\end{split}
\end{equation}
which may be understood as gauge invariance of~\eq{gauge field composite}. Motivated by such formulas, let us consider the following Schwinger functions:
\begin{equation} 
\begin{split}
\label{gauge invariant schwinger}
&\mathfrak{S}_{\mathfrak{I}}\bigl(\undertilde{x}_1, \cdots, \undertilde{x}_{m+l} \bigr) := \Bigl \langle \Omega, \bigl \lvert \Psi \bigr \rvert^2_{\nu_{r_1}  \nu_{\overline{r_1}}'}(\undertilde{x}_1 ) \cdots \bigl \lvert \Psi \bigr \rvert^2_{\nu_{r_m}  \nu_{\overline{r_m}}'}(\undertilde{x}_m )  F^2_{iji'j'}(\undertilde{x}_{m+1}) \cdots  F^2_{iji'j'}(\undertilde{x}_{m+l}) \Omega \Bigr \rangle \\
& \text{with } \mathfrak{I} \text{ given here as } \bigl\{ (\nu_{r_1}, \nu'_{\overline{r_1}}), \cdots,(\nu_{r_m}, \nu'_{\overline{r_m}}), (i_1,j_1,i'_1,j'_1), \cdots, (i_l,j_l,i'_l,j'_l)\bigr\}
\end{split}
\end{equation}
or again any permutation of the field operators and corresponding indices, which we check heuristically to be gauge-invariant.\footnote{\cite[Section 6]{FroMorStr81} investigates how a collection of gauge-invariant fields can describe a theory completely. More specifically,~\cite[p.564 (6.1)]{FroMorStr81} presents a list of composite particles in the $SU(2)$ theory with the Higgs and fermion fields in the fundamental representation. The list there does not coincide with the fields that appear in~\eq{gauge invariant schwinger}. At least in lattice gauge theories, we believe that it is possible to obtain any other complete set of gauge-invariant fields from~\eq{gauge invariant schwinger} (and vice versa) by using the projections introduced in~\cite[Section 6.3]{FroMorStr81}. However, we have not been able to prove this conjecture yet. Moreover, a continuum theory would pose a further problem regarding the co-locating operations. For now, we proceed with gauge-invariant Schwinger functions of the form~\eq{gauge invariant schwinger} only. \label{xxy} } The aforementioned reflection positivity can finally be imposed on Schwinger functions of the form~\eq{gauge invariant schwinger}.

This may be the continuum version of previous works on lattice approximation such as~\cite[p.448 Theorem  2.1]{OstSei78}.  
With analytic continuation to the Minkowski spacetime as in~\cite{OstSch73, OstSch75}, such positivity of~\eq{gauge invariant schwinger} together with some modification of~\cite[Theorem 3.7]{StrWig64} implies that we may (re)construct an actual Hilbert space completion of a subspace of $H$ containing $\Omega$. The Hilbert space may be regarded as a space of physical states, which justifies the previous interpretation of $\Omega$ as a physical vacuum. Details of such reconstructions will be presented in the next paper.

The remaining crucial issue is how one can actually construct co-located products such as Eq.~$(\ref{matter field composite})$, Eq.~$(\ref{gauge field composite0})$ and Eq.~$(\ref{gauge field composite})$ at least at the level of vacuum expectation values. Motivated by the operator product expansion~\cite[Chapter 20]{Wei96} and rigorous construction of Wick products~\cite{BruFreKoh96}, we assume existence of certain \textit{counter-terms} that cancel out ultraviolet singularity of Schwinger functions at coinciding points in order for the limits of point-splitting regularization to exist as in~\cite[pp.646--674]{BruFreKoh96}. As a specific example of such construction for the matter fields, consider
\begin{equation}
\label{matter field example}
    \mathfrak{S}_{(\nu_{r}, \nu'_{\overline{r}})}(\undertilde{{x}})= \Bigl \langle \Omega, \bigl \lvert \Psi \bigr \rvert^2_{\nu_r  \nu_{\overline{r}}'} (\undertilde{x} ) \Omega \Bigr \rangle
\end{equation}
which is the vacuum expectation value of~\eq{matter field composite}.\footnote{If~\eq{matter field composite} were just the Wick product of (Euclidean) free fields, then~\eq{matter field example} would be identically zero. However, since interacting fields on $\mathbb{R}^4$ are assumed here, we believe that it is unlikely for~\eq{matter field example} to just vanish identically. Again, a detailed analysis of such properties comes with actual construction of a theory, and we only consider~\eq{matter field example} as an illustrative example of the general cases that will be presented in the next section. The same rationale applies to~\eq{gauge field example 1} and~\eq{gauge field example 2}.}


Let us start with Schwinger functions of the form
\[
S_{(k_{r},\nu_{r}) ,(k'_{\overline{r}},\nu'_{\overline{r}})}\bigl( \undertilde{x}, \undertilde{x'} \bigr) = \Bigl \langle \Omega, \Psi_{k_r \nu_r}(\undertilde{x}) \Psi_{k'_{\overline{r}}\nu'_{\overline{r}}}(\undertilde{x'}) \Omega \Bigr \rangle
\]
and introduce certain \textit{counterterms} which enables renormalized co-locating operations. More specifically,

 subject to the same local gauge action as~\eq{matter field local action example}

\begin{assump}
\label{renormalization assumption1}    
Let us assume that there exist counterterms
\begin{equation}
\label{counterterm example}
 C_{(k_{r},\nu_{r}) ,(k'_{\overline{r}},\nu'_{\overline{r}})}   
\end{equation}
belonging to $\mathscr{S}'(\mathbb{R}^{4\times 2})$ and satisfying the following renormalization properties:

\begin{itemize}
    \item 
    The limit\begin{equation}  
    \begin{split}
    \label{renormalization limit}
         &\lim\limits_{n \to \infty} \Bigl( S_{(k_{r},\nu_{r}) ,(k'_{\overline{r}},\nu'_{\overline{r}})}-C_{(k_{r},\nu_{r}) ,(k'_{\overline{r}},\nu'_{\overline{r}})}\Bigr) \bigl(   \Delta_{n, k_r k'_{\overline{r}}} (f) \bigr) \\
          \end{split}
\end{equation}
exists for all $f\in \mathscr{S}(\mathbb{R}^{4})$ and any choice of $\bigl\{  \Delta_{n,k_r k'_{\overline{r}}}\bigr\}$.

\item~\eq{renormalization limit} is independent of choice of regularizers in the sense that for each $f$, any choice of $\bigl\{  \Delta_{n,k_r k'_{\overline{r}}}\bigr\}$ gives the same limit.
\end{itemize}

On the counterterms~\eq{counterterm example}, we impose the local gauge action as~\eq{matter field local action example}.\footnote{Local gauge action on the Schwinger functions is \textit{a priori} in the sense that there is already a canonical form motivated by classical gauge theories. That is, we have adjusted such classical structures to the vacuum expectation values (which are again c-number quantities) of quantum field operators. On the other hand, we are not able to find a canonical gauge action to impose on the counterterms at the level of general axioms. Therefore, action on the counterterms is introduced \textit{by hand} to establish a consistent gauge structure on the co-located Schwinger functions and remains to be verified case by case for each theory, We nevertheless believe that the counterterms needed for co-location in every physically relevant model must have this form of local gauge action. The same rationale applies to Assumption \ref{renormalization assumption2} and \ref{renormalization assumption3} below.}
\end{assump}

Note that for any values of the spinor indices, the mapping
\[
f \to \Bigl( S_{(k_{r},\nu_{r}), (k'_{\overline{r}},\nu'_{\overline{r}})}-C_{(k_{r},\nu_{r}), (k'_{\overline{r}},\nu'_{\overline{r}})}\Bigr) \bigl(   \Delta_{n,k_r k'_{\overline{r}}} (f) \bigr)
\]
is a temperate distribution on $\mathbb{R}^{4}$ for each $n$. Hence, the weak$^*$ limit~\eq{renormalization limit} defines a temperate distribution on  $\mathbb{R}^{4}$~\cite[Theorem 2.7]{Rud91}. 

The composite operator~\eq{matter field composite}, and therefore the Schwinger function~\eq{matter field example} should be gauge-invariant, and we would like to have~\eq{renormalization limit} as a rigorous definition of~\eq{matter field example}. Therefore,
\begin{invaria}
    We have to show that the limit~\eq{renormalization limit} is invariant under the gauge action in the sense that
    \begin{equation}
    \begin{split}
    \label{renormalization limit invariant}
       &\lim\limits_{n \to \infty} \Bigl( \bigl[g\cdot S \bigr]_{(k_{r},\nu_{r}) ,(k'_{\overline{r}},\nu'_{\overline{r}})}-\bigl[g\cdot C\bigr]_{(k_{r},\nu_{r}) ,(k'_{\overline{r}},\nu'_{\overline{r}})} \Bigr)\bigl(   \Delta_{n,k_r k'_{\overline{r}}} (f) \bigr)     \\
       &=\lim\limits_{n \to \infty} \Bigl( S_{(k_{r},\nu_{r}) ,(k'_{\overline{r}},\nu'_{\overline{r}})}-C_{(k_{r},\nu_{r}) ,(k'_{\overline{r}},\nu'_{\overline{r}})}\Bigr) \bigl(   \Delta_{n,k_r k'_{\overline{r}}} (f) \bigr)
       \end{split}
\end{equation}
for every $f \in \mathscr{S}(\mathbb{R}^{4})$, $g \in \mathcal{G}$ and any choice of $\Delta_{n,k_r k'_{\overline{r}}}$. 
\end{invaria}
By construction, however, verification of~\eq{renormalization limit invariant} is just a straightforward computation. With this invariance property, it is sensible to take ~\eq{renormalization limit} as the mathematical definition of~\eq{matter field example}. Note that this construction may be viewed a generalization of~\cite[Corollary 4.6]{BryFroSei81} and~\cite[Definition 5.1, Definition 5.2 and Proposition 5.3]{BruFreKoh96}.

Co-located products for gauge fields are more involved, as we have to go through two steps. For simplicity, we will only present the case of 
\begin{equation}
    \label{gauge field example 1}
    \mathcal{S}_{(\alpha, i ,j), (\alpha',i'j')}\bigl(\undertilde{x}, \undertilde{x'}\bigr)=\Bigl \langle \Omega, F_{\alpha i j}(\undertilde{x}) F_{\alpha' i' j'}(\undertilde{x'})\Omega\Bigr \rangle
\end{equation}
and
\begin{equation}
    \label{gauge field example 2}
    \mathfrak{S}_{(i,j,i',j')}(\undertilde{x})=\Bigl\langle \Omega, F^2_{i j i' j'}(\undertilde{x}) \Omega\Bigr \rangle
\end{equation}
in this section. With~\eq{gauge field composite0} in mind, we start with the following ingredients:
\begin{align}
    & t_{\alpha} \otimes t_{\alpha'}\Bigl( \partial_i \partial_{i'} S_{(\alpha,j), (\alpha', j')}-\partial_i \partial_{j'} S_{(\alpha,j), (\alpha', i')}-\partial_j \partial_{i'} S_{(\alpha,i), (\alpha', j')}+\partial_j \partial_{j'} S_{(\alpha,i), (\alpha', i')}\Bigr)\bigl(f \otimes h) \label{name1} \\
    &   t_\alpha \otimes [t_{\beta'}, t_{\gamma'}]_{\otimes}\Bigl( \partial_i S_{(\alpha,j), (\beta',i'), (\gamma',j')}-\partial_j S_{(\alpha,i), (\beta',i'), (\gamma',j')} \Bigr)\bigl(f \otimes A'{\Delta'}_{n'}(h)\bigr) \label{name2} \\
    & [t_{\beta}, t_{\gamma}]_{\otimes} \otimes t_\alpha \Bigl( \partial_{i'} S_{ (\beta,i), (\gamma,j),(\alpha',j')}-\partial_{j'} S_{ (\beta,i), (\gamma,j),(\alpha',i')} \Bigr)\bigl( {A\Delta}_{n}(f) \otimes h\bigr) \label{name3} \\
    & [t_\beta,t_\gamma]_{\otimes} \otimes [t_{\beta'}, t_{\gamma'}]_{\otimes}S_{(\beta,i), (\gamma, j), (\beta',i'), (\gamma',j')}\bigl( A\Delta_n(f) \otimes A'\Delta'_{n'}(h)\bigr) \label{name4}
\end{align}
for $f, h \in \mathscr{S}(\mathbb{R}^4)$ and any choices of $\Delta_n$, $\Delta'_{n'}$, $A$ and $A'$ to construct~\eq{gauge field example 1}. 

In analogy to the matter fields, we introduce counterterms for the co-locating operations:

\begin{assump}
\label{renormalization assumption2}
Let us assume that there exist counterterms
\begin{equation}
    \label{counterterm for FF}
  C_{(\alpha,i)}, \, C_{(\alpha,i), (\beta,j)}, \, C_{(\alpha,i), \,(\beta,j), (\gamma,k)} \, \text{ and } \, C_{(\beta,i), (\gamma,j),(\beta',i'), (\gamma',j')}
\end{equation}
belonging to $\mathscr{S}'(\mathbb{R}^{4 })$, $\mathscr{S}'(\mathbb{R}^{4 \times 2})$, $\mathscr{S}'(\mathbb{R}^{4 \times 3})$ and $\mathscr{S}'(\mathbb{R}^{4 \times 4})$ respectively and satisfying the following properties:
\begin{itemize}
    \item Symmetry under the commutation of index pairs and corresponding spacetime arguments in analogy to~\cite[(E3) in p.103]{OstSch73} is assumed for all individual Schwinger functions appearing in above equations (\ref{name1}), (\ref{name2}), (\ref{name3}), (\ref{name4}) as well as the counterterms in~\eq{counterterm for FF}. This is because the vector indices correspond to spin $1$ and therefore bosonic.

\item Schwinger functions for gauge fields subtracted by corresponding counterterms behave well under the co-locating operation with an additional spacetime argument in the sense that
\begin{equation}
\begin{split}
\label{sch colocating behavior0}
 \lim\limits_{n \to \infty}  \int_{(\mathbb{R}^4)^2} \bigl( S - C\bigr)_{(\alpha,i)}(\undertilde{x}) f(\undertilde{y}) \bigl[A\Delta_n(h) \bigr](\undertilde{x},\undertilde{y})  =  \bigl( S - C\bigr)_{(\alpha,i)}\bigl( A \bigl \lvert_{\text{diag}} f h  \bigr)
\end{split}
\end{equation}
\begin{equation}
\begin{split}
\label{sch colocating behavior}
     &\lim\limits_{n \to \infty}  \int_{(\mathbb{R}^4)^3} \bigl( S-C\bigr)_{(\alpha,i), (\beta, j)}(\undertilde{x}, \undertilde{y})  f(\undertilde{z})w(\undertilde{x}) \bigl[A\Delta_n(h) \bigr](\undertilde{y},\undertilde{z})  = \bigl( S-C\bigr)_{(\alpha,i), (\beta, j)}\Bigl( w \otimes A \bigl \lvert_{\text{diag}} fh \Bigr)  
\end{split}
\end{equation}
\begin{equation}
\begin{split}
\label{sch colocating behavior11}
     &\lim\limits_{n,n' \to \infty}  \int_{(\mathbb{R}^4)^4} \bigl( S-C\bigr)_{(\alpha,i), (\beta, j)}(\undertilde{x}, \undertilde{y})  f(\undertilde{x'})w(\undertilde{y'}) \bigl[A\Delta_n(h) \bigr](\undertilde{x},\undertilde{x'}) \bigl[A'\Delta'_{n'}(s) \bigr](\undertilde{y},\undertilde{y'})  \\
     &=  \bigl( S-C\bigr)_{(\alpha,i), (\beta, j)}\Bigl( A \bigl \lvert_{\text{diag}} fh \otimes  A' \bigl \lvert_{\text{diag}}ws \Bigr)
\end{split}
\end{equation}
and
\begin{equation}
\begin{split}
\label{sch colocating behavior1}
     &\lim\limits_{n \to \infty}  \int_{(\mathbb{R}^4)^4} f(\undertilde{x'})\bigl( S -C\bigr)_{(\alpha,i), (\beta, j),(\gamma,k)}(\undertilde{x}, \undertilde{y},\undertilde{z})  \bigl[A\Delta_n(h) \bigr](\undertilde{x},\undertilde{x'}) F(\undertilde{y},\undertilde{z}) \\
     &= \ \bigl( S -C\bigr)_{(\alpha,i), (\beta, j),(\gamma,k)}\Bigl( A \bigl \vert_{\text{diag}} fh \otimes F\Bigr)   
\end{split}
\end{equation}
for all $f, h, w, s \in \mathscr{S}(\mathbb{R}^4), F \in \mathscr{S}(\mathbb{R}^{4 \times 2})$,  $\Delta_n$, $A$ and indices. The integral notation is used to clarify additional spacetime arguments.

    \item The following co-locating limits within the spacetime arguments of Schwinger functions subtracted by corresponding counterterms
    \begin{equation}
        \label{FF limit 0}
       \lim\limits_{n \to \infty} \Bigl(  S_{(\alpha,i), (\beta,j)}- C_{ (\alpha,i), (\beta,j), }\Bigr)\bigl(    A\Delta_{n}(h)\bigr)
    \end{equation}    
    \begin{equation}
        \label{FF limit 1}
       \lim\limits_{n' \to \infty} \Bigl(  S_{(\alpha,j), (\beta',i'), (\gamma',j')}-  C_{ (\alpha,j), (\beta',i'), (\gamma',j')}\Bigr)\bigl( f \otimes  A'\Delta'_{n'}(h)\bigr)
    \end{equation}
    and
        \begin{equation}
        \label{FF limit 2}
       \lim\limits_{n,n' \to \infty} \Bigl( S_{(\beta,i), (\gamma, j), (\beta',i'), (\gamma',j')} - C_{(\beta,i), (\gamma, j), (\beta',i'), (\gamma',j')}\Bigr)\bigl( A \Delta_n(f) \otimes A'\Delta'_{n'}(h)\bigr)
    \end{equation}
    exist for all $f,h \in \mathscr{S}(\mathbb{R}^4)$, $A$, $A'$ and indices independently of choice of $\Delta_{n}$ and $\Delta'_{n'}$. They depend on $A$ and $A'$ only through $A \bigl \lvert_{\text{diag}}$ and $A' \bigl \lvert_{\text{diag}}$ in the sense that if we denote~\eq{FF limit 1} with $A'=1$ as $T_{\alpha \beta' \gamma'}(f \otimes h)$ and~\eq{FF limit 1} with general $A'$ as $T^{A'}_{\alpha \beta' \gamma'}(f \otimes h)$, then the following relation holds:
    \[
    T^{A'}_{\alpha \beta' \gamma'}(f \otimes h) = T_{\alpha \beta' \gamma'}\bigl(f \otimes A'\bigl\lvert_{\text{diag}}h\bigr)
    \]
    where we have suppressed the vector indices for brevity. Similar requirements are imposed on~\eq{FF limit 0} and~\eq{FF limit 2}.\footnote{As bilinear mappings of $(f,h)$,~\eq{FF limit 1} and (\ref{FF limit 2}) are separately continuous on $\mathscr{S}(\mathbb{R}^4) \times \mathscr{S}(\mathbb{R}^4)$ according to~\cite[p.46 Theorem 2.8]{Rud91} applied to each argument. Then, the nuclear theorem~\cite[p.141 Theorem V.12]{ReeSim72} implies that each of~\eq{FF limit 1} and (\ref{FF limit 2}) defines a unique temperate distribution on $\mathbb{R}^{4 \times 2}$ respectively. $T^{A'}_{\alpha \beta' \gamma'}$ is an element of $\mathscr{S}'(\mathbb{R}^{4 \times 2})$ obtained as such.} \footnote{Such conditions may look a bit unnatural. Nevertheless, together~\eq{sch colocating behavior}, they turn out to be crucial for establishing consistency of the gauge structures on co-located Schwinger functions (or composite field operators), as stated in~\eq{FF limit consistency 1}, (\ref{FF limit consistency 2}), (\ref{F2 gauge11}) and (\ref{F2 gauge invariance rigor}). Moreover, we believe that verification of such conditions may not be an impossible task when constructing an actual theory.} Moreover,~\eq{FF limit 1} is compatible with~\eq{sch colocating behavior1} in the sense that
    \begin{equation}
        \begin{split}
            \label{odd number pair}
           &\lim\limits_{n,n' \to \infty}  \int_{(\mathbb{R}^4)^4} f(\undertilde{x})\Bigl(  S_{(\alpha,j), (\beta',i'), (\gamma',j')}-  C_{ (\alpha,j), (\beta',i'), (\gamma',j')}\Bigr)(\undertilde{y}, \undertilde{x'},\undertilde{y'}) \bigl[A\Delta_n(h) \bigr](\undertilde{x},\undertilde{y})\bigl[A'\Delta'_{n'}(w) \bigr](\undertilde{x'},\undertilde{y'})  \\
     &=  \lim\limits_{n' \to \infty} \Bigl(  S_{(\alpha,j), (\beta',i'), (\gamma',j')}-  C_{ (\alpha,j), (\beta',i'), (\gamma',j')}\Bigr)\bigl( A\bigl\lvert_{\text{diag}} fh \otimes  A'\Delta'_{n'}(w)\bigr).
        \end{split}
    \end{equation}
    \end{itemize}
     We define the local gauge actions on counterterms in the same way as the Schwinger functions for gauge fields. For example,
   \begin{equation}
    \begin{split}
        \label{counterterm gauge action example}
         &t_\alpha \otimes t_{\beta} \bigl[ g \cdot C \bigr]_{ (\alpha, i), (\beta,j)}( f, h)\\
         &:=C_{(\alpha, i), (\alpha', i')}\Bigl( f gt_{\alpha}g^{-1} \otimes h gt_{\alpha'} g^{-1} \Bigr)-C_{(\alpha, i)}\Bigl(f gt_\alpha g^{-1}\Bigr)\otimes \Biggl(\int_{\mathbb{R}^4} h (\partial_{i'} g)g^{-1} \Biggr) \\
    &-\Biggl(\int_{\mathbb{R}^4} f (\partial_{i} g)g^{-1}\Biggr) \otimes C_{(\alpha', i')}\Bigl(h gt_{\alpha'} g^{-1}\Bigr)+\Biggl(\int_{\mathbb{R}^4} f (\partial_{i} g)g^{-1}\Biggr)\otimes\Biggl(\int_{\mathbb{R}^4} h (\partial_{i'} g)g^{-1}\Biggr)
   \end{split}
    \end{equation}
which is the same form as~\eq{gauge field local action example}.
\end{assump}

Before proceeding on, we need to check consistency of the local gauge action:
\begin{consis}
\label{consistency check1}
For any two (gauge-fixed) representatives in a gauge orbit, there exists by definition some $g \in \mathcal{G}$ that connects them by the local gauge action. Therefore, the following properties must be verified to show that Assumption~\ref{renormalization assumption1} and~\ref{renormalization assumption2} are consistent with the gauge structure.
\begin{enumerate}
  
      \item\label{1 claim} $\bigl(g \cdot S- g \cdot C\bigr)_{(\alpha,i)}$, $\bigl(g \cdot S- g \cdot C\bigr)_{(\alpha,i),(\beta,j)}$ and $\bigl(g \cdot S-g \cdot C\bigr)_{(\alpha,i),(\beta,j),(\gamma,k)}$ satisfy the same properties as~\eq{sch colocating behavior0}, (\ref{sch colocating behavior}), (\ref{sch colocating behavior11}) and (\ref{sch colocating behavior1}) respectively.

    \item\label{2 claim} The limits
\begin{equation}
        \label{FF limit consistency 0}
       \lim\limits_{n \to \infty} \Bigl(  \bigl[ g \cdot  S \bigr]_{(\alpha,i), (\beta,j)}- \bigl[ g \cdot  C \bigr]_{ (\alpha,i), (\beta,j), }\Bigr)\bigl(    A\Delta_{n}(h)\bigr)
    \end{equation}       
    \begin{equation}
        \label{FF limit consistency 1}
       \lim\limits_{n' \to \infty} \Bigl(   \bigl[ g \cdot  S \bigr]_{(\alpha,j), (\beta',i'), (\gamma',j')} - \bigl[g \cdot C\bigr]_{ (\alpha,j), (\beta',i'), (\gamma',j')}\Bigr)\bigl( f \otimes  A'\Delta'_{n'}(h)\bigr)
    \end{equation}
and
 \begin{equation}
        \label{FF limit consistency 2}
       \lim\limits_{n,n' \to \infty} \Bigl( \bigl[ g \cdot  S \bigr]_{(\beta,i), (\gamma, j), (\beta',i'), (\gamma',j')} - \bigl[ g \cdot  C \bigr]_{(\beta,i), (\gamma, j), (\beta',i'), (\gamma',j')}\Bigr)\bigl( A \Delta_n(f) \otimes A'\Delta'_{n'}(h)\bigr)
    \end{equation}
exist in the same way as~\eq{FF limit 0}, (\ref{FF limit 1}) and (\ref{FF limit 2}) respectively, for any $g \in \mathcal{G}$. That is, for each choice of $g$, the limits~\eq{FF limit consistency 1} and~\eq{FF limit consistency 2} exist for all $f,h \in \mathscr{S}(\mathbb{R}^4)$, $A$, $A'$ and indices independently of choice of $\Delta_{n}$ and $\Delta'_{n'}$. Moreover, they depend on $A$ and $A'$ only through $A \bigl \lvert_{\text{diag}}$ and $A' \bigl \lvert_{\text{diag}}$. Moreover,~\eq{FF limit consistency 1} satisfies the compatibility condition in the same form as~\eq{odd number pair}.
\end{enumerate}
\end{consis}
For verification of these properties, see Appendix \ref{appendix:a}.

With such consistency results in mind,~\eq{gauge field example 1} may be defined as follows:
\begin{align*}
       & t_\alpha \otimes t_{\alpha'}\mathcal{S}_{(\alpha,i,j)(\alpha',i',j')}(f \otimes h):= \label{FF def} \tag{FF}\\
    &t_{\alpha} \otimes t_{\alpha'}\Bigl( \partial_i \partial_{i'} S_{(\alpha,j), (\alpha', j')}-\partial_i \partial_{j'} S_{(\alpha,j), (\alpha', i')}-\partial_j \partial_{i'} S_{(\alpha,i), (\alpha', j')}+\partial_j \partial_{j'} S_{(\alpha,i), (\alpha', i')}\Bigr)\bigl(f \otimes h) \label{name1'} \tag{FF-1}\\
   &+  t_\alpha \otimes [t_{\beta'}, t_{\gamma'}]\lim\limits_{n' \to \infty}\partial_i\bigl(  S- C\bigr)_{(\alpha,j), (\beta',i'), (\gamma',j')}\bigl(f \otimes {\Delta'}_{n'}(h)\bigr) \label{name2'} \tag{FF-2-i}\\
   &-  t_\alpha \otimes [t_{\beta'}, t_{\gamma'}]\lim\limits_{n' \to \infty}\partial_j\bigl(  S - C \bigr)_{(\alpha,i), (\beta',i'), (\gamma',j')}\bigl(f \otimes {\Delta'}_{n'}(h)\bigr) \label{name22'} \tag{FF-2-ii}\\
   &+ [t_{\beta}, t_{\gamma}] \otimes t_\alpha \lim\limits_{n \to \infty}\partial_{i'} \bigl( S- C\bigr)_{ (\beta,i), (\gamma,j),(\alpha',j')}\bigl( {\Delta}_{n}(f) \otimes h\bigr) \label{name3'} \tag{FF-3-i}\\
   &- [t_{\beta}, t_{\gamma}] \otimes t_\alpha \lim\limits_{n \to \infty}\partial_{j'}\bigl(  S - C\bigr)_{ (\beta,i), (\gamma,j),(\alpha',i')}\bigl( {\Delta}_{n}(f) \otimes h\bigr) \label{name33'} \tag{FF-3-ii} \\
   &+[t_\beta,t_\gamma] \otimes [t_{\beta'}, t_{\gamma'}]\lim\limits_{n,n' \to \infty}\bigl( S- C\bigr)_{(\beta,i), (\gamma, j), (\beta',i'), (\gamma',j')} \bigl( \Delta_n(f) \otimes \Delta'_{n'}(h)\bigr) \label{name4'} \tag{FF-4}
 \end{align*}
 where the limits in (\ref{name2'}), (\ref{name22'}), (\ref{name3'}) and (\ref{name33'}) exist due to~\eq{FF limit 1} combined with the symmetry assumption and the limit in (\ref{name4'}) due to~\eq{FF limit 2}.

Here, consistency checks under the local gauge structure as in~\eq{FF limit consistency 1} and (\ref{FF limit consistency 2}) imply the same consistency for ~\eq{FF def}. In fact, one observes that the following local gauge action
\begin{equation}
\label{F2 non colocate gauge action}
   t_{\alpha} \otimes t_{\alpha'}\bigl[ g \cdot \mathcal{S} \bigr]_{(\alpha,i,j),(\alpha',i',j')}\bigl( f \otimes h \bigr)=\mathcal{S}_{(\alpha,i,j),(\alpha',i',j')}\bigl( f gt_\alpha g^{-1} \otimes h g t_{\alpha'} g^{-1} \bigr)
\end{equation}
is derived from local gauge actions for Schwinger functions corresponding to gauge potentials as in~\eq{gauge field local action example} and counterterms as in~\eq{counterterm gauge action1} and (\ref{counterterm gauge action2}) combined with the co-locating operations in (\ref{name2'}, \ref{name3'}, \ref{name4'}). 

Actual verification is a bit lengthy but straightforward. Taking local gauge action on (\ref{name1}, \ref{name2}, \ref{name3}, \ref{name4}) leads to
\begin{equation}
\begin{split}
&t_{\alpha} \otimes t_{\alpha'} \Biggl\{g \cdot\Bigl( \partial_i \partial_{i'} S_{(\alpha,j), (\alpha', j')}-\partial_i \partial_{j'} S_{(\alpha,j), (\alpha', i')}-\partial_j \partial_{i'} S_{(\alpha,i), (\alpha', j')}+\partial_j \partial_{j'} S_{(\alpha,i), (\alpha', i')}\Bigr)\Biggr\}\bigl(f \otimes h)\\
&=\Bigl( \partial_i \partial_{i'} S_{(\alpha,j), (\alpha', j')}-\partial_i \partial_{j'} S_{(\alpha,j), (\alpha', i')}-\partial_j \partial_{i'} S_{(\alpha,i), (\alpha', j')}+\partial_j \partial_{j'} S_{(\alpha,i), (\alpha', i')}\Bigr)\bigl( f gt_\alpha g^{-1} \otimes h gt_{\alpha'}g^{-1} \bigr)
\end{split}
\end{equation}

\begin{equation}
\begin{split}
\label{name2''}
   &t_\alpha \otimes [t_{\beta'}, t_{\gamma'}]_\otimes\Biggl\{g \cdot \Bigl( \partial_i S_{(\alpha,j), (\beta',i'), (\gamma',j')}-\partial_j S_{(\alpha,i), (\beta',i'), (\gamma',j')} \Bigr)\Biggr\}\bigl(f \otimes {\Delta'}_{n'}(h)\bigr)\\
   &=\Bigl( \partial_{i'} S_{ (\beta,i), (\gamma,j),(\alpha',j')}-\partial_{j'} S_{ (\beta,i), (\gamma,j),(\alpha',i')} \Bigr) \bigl(f gt_\alpha g^{-1} \otimes  {\Delta'}_{n'}(h)[gt_{\beta'}g^{-1},gt_{\gamma'}g^{-1}]\bigr)
\end{split}
\end{equation}

\begin{equation}
\begin{split}
[t_{\beta}, t_{\gamma}]_\otimes \otimes t_\alpha \Biggl\{ g \cdot \Bigl( \partial_{i'} S_{ (\beta,i), (\gamma,j),(\alpha',j')}-\partial_{j'} S_{ (\beta,i), (\gamma,j),(\alpha',i')} \Bigr) \Biggr\}\bigl( {\Delta}_{n}(f) \otimes h\bigr)
\end{split}   
\end{equation}
and
\begin{equation}
\begin{split}
[t_\beta,t_\gamma]_\otimes \otimes [t_{\beta'}, t_{\gamma'}]_\otimes\Bigl( g\cdot S_{(\beta,i), (\gamma, j), (\beta',i'), (\gamma',j')}\Bigr)\bigl( \Delta_n(f) \otimes \Delta'_{n'}(h)\bigr)
\end{split}   
\end{equation}
where 
\[
\Bigl({\Delta}_{n}(f)[gt_{\beta}g^{-1},gt_{\gamma}g^{-1}]\Bigr)(\undertilde{x},\undertilde{y}):= {\Delta}_{n}(f)(\undertilde{x},\undertilde{y})\bigl[g(\undertilde{x})t_{\beta}g^{-1}(\undertilde{x}),g(\undertilde{y})t_{\gamma}g^{-1}(\undertilde{y}) \bigr]
\]
and similarly for $ {\Delta'}_{n'}(h)[gt_{\beta'}g^{-1},gt_{\gamma'}g^{-1}]$. 

Observe that the LHS of~\eq{name2''} can be written as
\[
\Biggl\{g \cdot \Bigl( \partial_i S_{(\alpha,j), (\beta',i'), (\gamma',j')}-\partial_j S_{(\alpha,i), (\beta',i'), (\gamma',j')} \Bigr)\Biggr\}\Bigl(ft_\alpha \otimes {\Delta'}_{n'}(h)[t_{\beta'}, t_{\gamma'}]_\otimes\Bigr)
\]
and that $I \otimes \mathscr{P}$ commutes with the temperate distributions. Accordingly, taking $I \otimes \mathscr{P}$ on the both sides of~\eq{name2''} leads to

Finally, we construct~\eq{gauge field example 2} from~\eq{gauge field example 1} in a similar way. 

\begin{assump}
\label{renormalization assumption3}
Let us assume that there exist counterterms
\begin{equation}
    \mathcal{C}_{(\alpha,i,j)(\alpha',i',j')}(\undertilde{x},\undertilde{x'}) 
\end{equation}
belonging to $\mathscr{S}'(\mathbb{R}^{4 \times 2})$ such that the limit
\begin{equation}
    \begin{split}
    \label{F2 rigorous def}
    \lim\limits_{n \to \infty} \Bigl(\mathcal{S}_{(\alpha,i,j)(\alpha',i',j')} -  \mathcal{C}_{(\alpha,i,j)(\alpha',i',j')}\Bigr)\bigl(A\Delta_{n}(f)\bigr)   
    \end{split}
\end{equation}
exists for all $f \in \mathscr{S}(\mathbb{R}^4)$, $A$ and indices independently of choice of $\Delta_{n}$. Moreover, it depends on $A$ only through $A \bigl \lvert_{\text{diag}}$ in the same sense as~\eq{FF limit 1} or (\ref{FF limit 2}). We impose on the counterterms a local gauge action of the form
\begin{equation}
    \label{counterterm2 gauge action}
   t_{\alpha} \otimes t_{\alpha'}\bigl( g \cdot \mathcal{C} \bigr)_{(\alpha,i,j)(\alpha',i',j')}(\undertilde{x},\undertilde{x'}):=\mathcal{C}_{(\alpha,i,j)(\alpha',i',j')}(\undertilde{x},\undertilde{x'})g(\undertilde{x})t_\alpha g^{-1}(\undertilde{x}) \otimes g(\undertilde{x'})t_{\alpha'} g^{-1}(\undertilde{x'})
\end{equation}
which, of course, should be understood in the sense of temperate distributions.
\end{assump}
Just like~\eq{FF limit consistency 1} and (\ref{FF limit consistency 2}), we must check that~\eq{F2 rigorous def} leads to a consistent gauge structure as well as the invariance property computed heuristically with field operators in~\eq{F2 gauge invariance heuristic}. More specifically, the following two claims must be verified:
\begin{consis}
     The limit
    \begin{equation}
    \begin{split}
    \label{F2 gauge11}
    \lim\limits_{n \to \infty} \Bigl(\bigl[g \cdot \mathcal{S}\bigr]_{(\alpha,i,j)(\alpha',i',j')} -  \bigl[ g \cdot \mathcal{C}\bigr]_{(\alpha,i,j)(\alpha',i',j')}\Bigr)\bigl(A\Delta_{n}(f)\bigr)   
    \end{split}
\end{equation}
exists in the same way as~\eq{F2 rigorous def} for each $g \in \mathcal{G}$.
\end{consis}

\begin{invaria}
     Such limits are invariant under the gauge action in the sense that
\begin{equation}
    \begin{split}
    \label{F2 gauge invariance rigor}
    &\lim\limits_{n \to \infty} \Bigl(\bigl[g \cdot \mathcal{S}\bigr]_{(\alpha,i,j)(\alpha,i',j')} -  \bigl[ g \cdot \mathcal{C}\bigr]_{(\alpha,i,j)(\alpha,i',j')}\Bigr)\bigl(A\Delta_{n}(f)\bigr)   \\
    &= \lim\limits_{n \to \infty} \Bigl(\mathcal{S}_{(\alpha,i,j)(\alpha,i',j')} -   \mathcal{C}_{(\alpha,i,j)(\alpha,i',j')}\Bigr)\bigl(A\Delta_{n}(f)\bigr) 
    \end{split}
\end{equation}
for all $g \in \mathcal{G}$. Note that gauge indices are contracted here.
\end{invaria}

Both are straightfoward computations, which we leave to Appendix \ref{appendix:a}.

~\eq{F2 non colocate gauge action} and~\eq{counterterm2 gauge action} imply that
\begin{equation}
\begin{split}
\label{FF invariance proof1}
&\Bigl(\bigl[g \cdot \mathcal{S}\bigr]_{(\alpha,i,j)(\alpha',i',j')} -  \bigl[ g \cdot \mathcal{C}\bigr]_{(\alpha,i,j)(\alpha',i',j')}\Bigr)\bigl(A\Delta_{n}(f)\bigr)\\
&=\Bigl(\mathcal{S}_{(\beta,i,j)(\beta',i',j')}-\mathcal{C}_{(\beta,i,j)(\beta',i',j')}\Bigr)\Bigl(\Delta_n(f) \Tr\bigl(gt_\beta g^{-1} t_\alpha\bigr) \otimes \Tr\bigl(gt_{\beta'} g^{-1} t_{\alpha'}\bigr) A\Bigr)
\end{split}
\end{equation}
where the test function in the second line of~\eq{FF invariance proof1} is understood as
\[
\Delta_n(f)(\undertilde{x},\undertilde{x'}) \Tr\bigl(g(\undertilde{x})t_\beta g^{-1}(\undertilde{x}) t_\alpha\bigr)\Tr\bigl(g(\undertilde{x'})t_{\beta'} g^{-1}(\undertilde{x'})t_{\alpha'}\bigr) A(\undertilde{x},\undertilde{x'}) \in \mathscr{S}(\mathbb{R}^{4 \times 2}).
\]
It is clear that  
\[
A^g_{ \beta  \alpha \beta' \alpha'}(\undertilde{x},\undertilde{x'}):=\Tr\bigl(g(\undertilde{x})t_\beta g^{-1}(\undertilde{x}) t_\alpha \bigr)\Tr\bigl(g(\undertilde{x'})t_{\beta'} g^{-1}(\undertilde{x'})t_{\alpha'}\bigr) 
\]
satisfies Definition~\ref{gauge subtlety} for all values of the indices, and so does $A^g_{ \beta  \alpha \beta' \alpha'}A$. Therefore,~\eq{F2 gauge11} exists for all $f$, $A$, indices independently of choice of $\Delta_n$. To check dependence on $A$, let us introduce the following notations:
\begin{itemize}
    \item $F_{\alpha \alpha'}(f)$ for~\eq{F2 rigorous def} with $A=1$
    \item $F^A_{\alpha \alpha'}(f)$ for~\eq{F2 rigorous def} with general $A$
    \item ${\leftindex^g{F}}_{\alpha \alpha'}(f)$ for~\eq{F2 gauge11} with $A=1$
    \item ${\leftindex^g{F}}^{A}_{\alpha \alpha'}(f)$ for~\eq{F2 gauge11} with general $A$
\end{itemize}
where the vector indices are suppressed for brevity. Then, by assumption, we have 
\begin{equation}
\label{consistency}
    F^A_{\alpha \alpha'}(f)=F_{\alpha \alpha'}\bigl( A \bigl\lvert_{\text{diag}}f \bigr)
\end{equation}
which must be true for $F^g_{\alpha \alpha'}$ and $F^{g,A}_{\alpha \alpha'}$ as well. Indeed, combination of~\eq{FF invariance proof1} and~\eq{consistency} yields
\[
{\leftindex^g{F}}^A_{\alpha \alpha'}(f)=F_{\beta \beta'}^{A^g_{ \beta  \alpha \beta' \alpha'}A}(f)=F_{\beta \beta'}\bigl(  A^g_{ \beta  \alpha \beta' \alpha'}\bigl\lvert_{\text{diag}} A\bigl\lvert_{\text{diag}} f \bigr) 
\]
while
\[
{\leftindex^g{F}}_{\alpha \alpha'}(f)=F_{\beta \beta'}^{A^g_{ \beta  \alpha \beta' \alpha'}}(f)=F_{\beta \beta'}\bigl( A^g_{ \beta  \alpha \beta' \alpha'}\bigl\lvert_{\text{diag}}f \bigr)
\]
so that
\[
{\leftindex^g{F}}^A_{\alpha \alpha'}(f)={\leftindex^g{F}}_{\alpha \alpha'}\bigl( A \bigl\lvert_{\text{diag}}f \bigr)
\]
which is the desired result.

For the second claim, taking $\Tr_{\otimes}$ defined by~\eq{noncolocated trace} on both sides of~\eq{F2 non colocate gauge action} and~\eq{counterterm2 gauge action} smeared by $\Delta_n(f)A$ leads to
\begin{equation}
\begin{split}
\label{FF invariance 2}
 &\bigl[g \cdot \mathcal{S}- g \cdot \mathcal{C}\bigr]_{(\alpha,i,j)(\alpha',i',j')}\bigl(\Delta_{n}(f) \delta_{\alpha \alpha'}A\bigr)\\
 &=\bigl[\mathcal{S}- \mathcal{C}\bigr]_{(\alpha,i,j),(\alpha',i',j')}\Bigl( \Delta_n(f)\Tr_{\otimes}\bigl( \Delta_n(f) gt_\alpha g^{-1} \otimes  g t_{\alpha'} g^{-1} \bigr) A\Bigr)\\
 &=\bigl[\mathcal{S}- \mathcal{C}\bigr]_{(\alpha,i,j),(\alpha',i',j')}\Bigl(  \Delta_n(f)(\undertilde{x},\undertilde{x'}) \Tr\bigl(g(\undertilde{x})t_\alpha g^{-1}(\undertilde{x})   g(\undertilde{x'}) t_{\alpha'} g^{-1}(\undertilde{x'}) \bigr)A(\undertilde{x},\undertilde{x'})\Bigr)
\end{split}
\end{equation}
where we have used linearity to put $\Tr_{\otimes}$ into temperate distributions in the second line and explicitly displayed the arguments of test functions in the third line for clarity. Now, for any $g \in \mathcal{G}$, the function
\[
\Tr\bigl(g(\undertilde{x})t_\alpha g^{-1}(\undertilde{x})   g(\undertilde{x'}) t_{\alpha'} g^{-1}(\undertilde{x'}) \bigr) A(\undertilde{x},\undertilde{x'})
\]
again satisfies Definition~\ref{gauge subtlety} and is equal to $\delta_{\alpha \alpha'}A \bigl \lvert_{\text{diag}}$ when restricted to the diagonal. Therefore, the limit of~\eq{FF invariance 2} as $n \to \infty$ is identical to that of
\[
\bigl[\mathcal{S}- \mathcal{C}\bigr]_{(\alpha,i,j),(\alpha,i',j')}\Bigl(  A\Delta_n(f)\Bigr)=\bigl[\mathcal{S}- \mathcal{C}\bigr]_{(\alpha,i,j),(\alpha',i',j')}\Bigl(  \delta_{\alpha \alpha'}A\Delta_n(f)\Bigr)
\]
for all $g \in \mathcal{G}$, which proves the second claim. Therefore, we may regard (the second line of)~\eq{F2 gauge invariance rigor} with $A=1$ as the definition of~\eq{gauge field example 2} smeared by $f$.

\section{The Axioms for Schwinger Functions of a Quantum Gauge Theory}
\label{local axioms}

In this section, we formally state the OS axioms adjusted to encompass non-Abelian gauge symmetry, whose motivations and simple examples are presented in the previous section:

\begin{definition}
\label{defsch}
    The following five collections of temperate distributions
   \[
    \cup_{m \in \mathbb{N} \cup \{0\} }\Bigl\{ S_{ I}  \, \bigl \lvert \,  S_I \in \mathscr{S}'(\mathbb{R}^{4m}) \text{ and } I \text{ runs through all possible cases for } m \text{ elements} \Bigr\} \tag{S1} \label{schwinger def1}
   \]

     \[
          \cup_{m \in \mathbb{N} \cup \{0\}}\Bigl\{ \mathcal{S}_{ \mathcal{I}}  \, \bigl \lvert \,  \mathcal{S}_\mathcal{I} \in \mathscr{S}'(\mathbb{R}^{4m})\text{ and } \mathcal{I} \text{ runs through all possible cases for } m \text{ elements} \Bigr\} \tag{S2} \label{schwinger def2}
     \]

    \[
   \cup_{m \in \mathbb{N} \cup \{0\}} \Bigl\{ \mathfrak{S}_{\mathfrak{I}}   \, \bigl \lvert \,  \mathfrak{S}_{\mathfrak{I}} \in \mathscr{S}'(\mathbb{R}^{4m}) \text{ and } \mathfrak{I} \text{ runs through all possible cases for } m \text{ elements}\Bigr\} \tag{S3} \label{schwinger def3}
    \]
with the index sets as in Definition~\ref{index and permutation} and $S_{\emptyset}=\mathcal{S}_{\emptyset}=\mathfrak{S}_{\emptyset}=1$ for $m=0$ are called the Schwinger functions (under a gauge fixing) for a quantum Yang-Mills theory with the gauge group $G$ and the following two collections of temperate distributions   
     \begin{equation}
     \tag{C1} \label{counter def1}
         \begin{aligned}
         &\cup_{m \in \mathbb{N} }\Bigl\{ C_{ I}  \, \bigl \lvert \,  C_I \in \mathscr{S}'(\mathbb{R}^{4 \times 2m}) \text{ and }I \text{ consists of } m \text{ adjacent pairs of the forms } \\
              &(k_r, \nu_r) \, \& \,(k'_{\overline{r}}, \nu'_{\overline{r}}) \text{ and }(\alpha,i) \, \& \, (\alpha',i') \text{ so that it has } 2m \text{ elements in total.}\Bigr\}     \end{aligned}
     \end{equation}
       
     \[
         \cup_{m \in \mathbb{N}}\Bigl\{ \mathcal{C}_{(\alpha_1,i_1,j_1),(\alpha'_1,i'_1,j'_1), \cdots ,(\alpha_m,i_m,j_m),(\alpha'_m,i'_m,j'_m) }  \, \bigl \lvert \,  \mathcal{C}_{(\alpha_1,i_1,j_1),(\alpha'_1,i'_1,j'_1), \cdots ,(\alpha_m,i_m,j_m),(\alpha'_m,i'_m,j'_m) } \in \mathscr{S}'(\mathbb{R}^{4 \times 2m})  \Bigr\} \tag{C2} \label{counter def2}
    \]
 are called the corresponding counterterms\footnote{Note that the index sets appearing in~\eq{counter def1} and (\ref{counter def2}) are special cases of $I$ and $\mathcal{I}$ respectively. We specified indices explicitly in~\eq{counter def2} rather than using the notation $\mathfrak{I}$ for the sake of clarity.} for above Schwinger functions if they satisfy the following axioms:

\begin{itemize}
\item[$\blacksquare$] For~\eq{schwinger def1}, $(\ref{counter def1})$ and $(\ref{counter def2})$,

\begin{itemize}[label=$\diamond$]
   
   \item Euclidean Covariance

This applies to the spinor indices only and may be formulated in the same way as~\cite[p.102, (E1)]{OstSch73}. Note that for $\mathcal{I}$ or $\mathfrak{I}$, the spinor indices in each tuple must be understood as tensor product of corresponding representations.

    \item (Anti)Commutation 

\[
S_{\sigma \cdot I}\bigl( \undertilde{x}_{\sigma(1)}, \cdots \undertilde{x}_{\sigma(m)}\bigr) = \pm S_{ I}\bigl( \undertilde{x}_{1}, \cdots \undertilde{x}_{m}\bigr)\tag{AC} \label{AC}
\]
and
\[
C_{\sigma \cdot I}\bigl( \undertilde{x}_{\sigma(1)}, \cdots \undertilde{x}_{\sigma(2m)}\bigr) = \pm C_{ I}\bigl( \undertilde{x}_{1}, \cdots \undertilde{x}_{2m}\bigr) \tag{AC'} \label{AC'}
\]
where $+$ is for the cases in which $\sigma$ carries out an even number of transpositions of fermionic indices, and $-$ is for an odd number of such transpositions. Note that~\eq{AC} and (\ref{AC'}) correspond to~\cite[p.103, (E3)]{OstSch73}.

~\eq{counter def2} is invariant under permutation of arguments and corresponding indices just like $+$ sign in~\eq{AC'}. However, these counterterms have anti-commutation properties for permutation of the spinor indices within each tuple, which corresponds to a single argument. That is, interchange of $i_l  \, \& \, j_l$ or $i'_l \, \& \, j'_l$ for any $l \in \{1,\cdots,m\}$ yields a minus sign.\footnote{For example, $\mathcal{C}_{(\alpha,i,j),(\alpha',i',j') }(\undertilde{x},\undertilde{x'})=\mathcal{C}_{(\alpha,j,i),(\alpha',j',i') }(\undertilde{x},\undertilde{x'})=-\mathcal{C}_{(\alpha,j,i),(\alpha',i',j') }(\undertilde{x},\undertilde{x'})=-\mathcal{C}_{(\alpha,i,j),(\alpha',j',i') }(\undertilde{x},\undertilde{x'})$ for $m=1$.}

\end{itemize}

\item[$\blacksquare$] For~\eq{schwinger def2} with respect to \eq{schwinger def1} and $(\ref{counter def1})$
\begin{itemize}[label=$\diamond$]
    \item Renormalized Co-location


    \[
\lim\limits_{x \to y} \tag{S2-1} \label{S21}
\]
\end{itemize}

\item[$\blacksquare$] For~\eq{schwinger def3} with respect to~\eq{schwinger def2} and $(\ref{counter def2})$
\begin{itemize}[label=$\diamond$]
    \item Renormalized Co-location

    \[
\lim\limits_{x \to y} \tag{CL1} \label{CL1}
\]
\end{itemize}

\item[$\blacksquare$] For all of~\eq{schwinger def1}, (\ref{schwinger def2}), (\ref{schwinger def3}), (\ref{counter def1}) and (\ref{counter def2})
\begin{itemize}[label=$\diamond$]

\item Growth Bounds 

    There exists a continuous semi-norm $\lVert \cdot \rVert$ on $\mathscr{S}(\mathbb{R}^4)$ and a constant $K>0$ such that
\[
 \max\Bigl\{  \bigl \lvert S_{I}\bigl( f_1 \otimes \cdots \otimes f_{m} \bigr) \bigr\rvert, \bigl \lvert \mathcal{S}_{\mathcal{I}}\bigl( f_1 \otimes \cdots \otimes f_{m} \bigr) \bigr\rvert, \bigl \lvert \mathfrak{S}_{\mathfrak{I}}\bigl( f_1 \otimes \cdots \otimes f_{m} \bigr) \bigr\rvert \Bigr\}\leq K^m m! \prod_{l=1}^{m}\lVert f_l \rVert \tag{GB1} \label{GB1}
\]
and another constant $L>0$ such that
\[
 \max\Bigl\{  \bigl \lvert C_{I}\bigl( f_1 \otimes \cdots \otimes f_{2m} \bigr) \bigr\rvert, \bigl \lvert \mathcal{C}_{\mathcal{I}}\bigl( f_1 \otimes \cdots \otimes f_{2m} \bigr) \bigr\rvert \Bigr\}\leq K^{2m} (2m)!^L \prod_{l=1}^{2m}\lVert f_l \rVert \tag{GB2} \label{GB2}
\]
for any choice of $m \in \mathbb{N}$, $f_1, \cdots, f_{m}, f_{m+1}, \cdots, f_{2m} \in \mathscr{S}(\mathbb{R}^4)$ and index sets for each type of Schwinger functions or couterterms.\footnote{(\ref{GB2}) is of the form stated in~\cite[p.98]{time79} or~\cite[p.287]{OstSch75}, which is sufficient for analytic continuation into the Minkowski metric. However,~\eq{GB1} is more restrictive in that $L=1$ there. The reason for choosing such bounds for Schwinger functions is to establish Euclidean path integral via the moment problem; see Section~\ref{squant} with Proposition~\ref{moment prob} in particular. Of course, Schwinger functions will also be analytically continued into the Minkowski metric, together with counterterms, to reconstruct the state space, culminating in a physical Hilbert space. }

\end{itemize}

\item[$\blacksquare$] For~\eq{schwinger def3}
\begin{itemize}[label=$\diamond$]
\item Reflection Positivity
For any $\uline{f}$ as in Definition~\ref{schwartz sequence def},
\[
\sum_{m,m'}  \mathfrak{S}_{  \mathfrak{I} \sqcup   \mathfrak{I}'} \Bigl(  [\Theta \uline{f}]_{m,\mathfrak{I}} \otimes \uline{f}_{m', \mathfrak{I}'} \Bigr) \geq 0. \tag{RP} \label{RP}
\]

  \item Cluster Decomposition\footnote{In~\cite[p.383 Corollary 4.6]{BryFroSei81}, gauge-invariant Schwinger functions are shown to obey all of the Euclidean axioms with cluster decomposition as a possible exception. Nevertheless, we believe the property to be true as well.}

Let $\uline{f}$ and $\uline{g}$ be as in Definition~\ref{schwartz sequence def}. Then, for any $4$-vector $a:=(0, \overrightarrow{a})$ with nonzero $\overrightarrow{a} \in \mathbb{R}^3$,
\[
\lim\limits_{\lambda \to \infty} \sum_{m,m'} \Biggr\{ \mathfrak{S}_{  \mathfrak{I} \sqcup   \mathfrak{I}'} \Bigl(  [\Theta \uline{f}]_{m,I} \otimes \uline{g}^{(\lambda a)}_{m', I' } \Bigr) -\mathfrak{S}_{  \mathfrak{I}}  \Bigl( [\Theta \uline{f}]_{m,I} \Bigr) \mathfrak{S}_{  \mathfrak{I}'} \Bigl(\uline{g}_{m', I'} \Bigr) \Biggl\}=0 \tag{CD} \label{CD}
\]
where $\uline{g}^{(\lambda a)}_{m', I'}$ is the translation of all arguments of $\uline{g}_{m', I'}$ by $\lambda a$ as defined in~\cite[p.87]{OstSch73}.

\end{itemize}

\end{itemize}
\end{definition}

\begin{remark}
From above axioms, we may deduce for~\eq{schwinger def2} and (\ref{schwinger def3}) Euclidean covariance as well as the following commutation properties:
\[
\mathcal{S}_{{\sigma \cdot \mathcal{I}}}\bigl( \undertilde{x}_{\sigma(1)}, \cdots \undertilde{x}_{\sigma(m)}\bigr) =  \mathcal{S}_{ \mathcal{I}}\bigl( \undertilde{x}_{1}, \cdots \undertilde{x}_{m}\bigr)
\]

\[
\mathfrak{S}_{{\sigma \cdot \mathfrak{I}}}\bigl( \undertilde{x}_{\sigma(1)}, \cdots \undertilde{x}_{\sigma(m)}\bigr) =  \mathfrak{S}_{ \mathfrak{I}}\bigl( \undertilde{x}_{1}, \cdots \undertilde{x}_{m}\bigr).
\]
Moreover, we derive (anti)commutation properties for permutation of the spinor indices within each tuple, which corresponds to a single argument in the temperate distribution. That is,
\begin{enumerate}
    
    \item In both $\mathcal{I}$ and $\mathfrak{I}$, interchange of $\nu_{r}$ and $\nu'_{\overline{r}}$ within any $(\nu_{r},\nu'_{\overline{r}})$ yields a minus sign if they are fermionic.
    
    \item In $\mathcal{I}$, interchange of $i$ and $j$ in any $(\alpha, i,j)$ yields a minus sign.

    \item In $\mathfrak{I}$, interchange of $i  \, \& \, j$ or $i' \, \& \, j'$ in any $(i,j,i',j')$ yields a minus sign. However, transposition of the two pairs does not cause any change in the sign. 
\end{enumerate}

\end{remark}

Such Schwinger functions and counterterms are designed to admit a local gauge structure, defined as follows:
\begin{definition}
\label{local action formal def}
    The local action of the gauge group $G$ on Schwinger functions in~\eq{schwinger def1} is defined on 
\end{definition}

It is not difficult to check via density argument that this action is indeed a group action:
\begin{prop}
\end{prop}
\begin{proof}
    dd
\end{proof}

With respect to the local gauge symmetry given by Definition~\ref{local action formal def}, we may establish gauge invariance of~\eq{schwinger def3} in the following sense, which justifies the requirement that~\eq{schwinger def3} must satisfy reflection positivity: 
\begin{prop}
\end{prop}
\begin{proof}
    dd
\end{proof}

\section{Sanity Check-1 : 2D Pure Yang-Mills Theories}

The main reason we only considered $\mathbb{R}^4$ so far is that failure of the spin-statistics theorem in lower dimensions makes the (anti)commutation axiom complicated. 

However, for a pure Yang-Mills theory in $2$ Euclidean dimension, we may still check if the above axioms scheme works properly. In fact, complete axial gauge makes the issue of gauge fixing quite simple. Moreover, the field strength tensor $F$ is a Gaussian field on the Lie group $G$, which implies that their co-located products may still be defined as the (Euclidean) Wick product.


\section{Sanity Check-2 : 3D Yang-Mills Higgs Model}
\label{squant}

Stochastic quantization (SQ) aims to construct the functional measure for a theory directly. My axiom scheme is related to SQ via the \textit{moment problem}. That is, the Schwinger functions are moments of the functional measure in question, as presented in~\cite{moment,moment2}. 

In this section, we reconstruct the 3D Yang-Mills-Higgs theory described in~\cite{stoc2}, starting from the gauge-invariant Schwinger functions $\mathfrak{S}_{\mathfrak{J}}$ (adjusted to 3D) by further imposing suitable dynamics as well as Nelson-Symanzik (NS) positivity.

First, we define what it precisely means by Nelson-Symanzik positivity:
\begin{definition}
    dd
\end{definition}

With the growth bounds and NS positivity, it is possible to construct a unique gauge-invariant Borel probability measure $\mu$ on $\mathscr{S}'(\mathbb{R}^4)$ having the Schwinger functions $\mathfrak{S}_{\mathfrak{J}}$ as moments:
\begin{prop}
\label{moment prob}
dd
\end{prop}
\begin{proof}
    (I owe the uniqueness part to Prof. Abdelmalek Abdesselam~\cite{MO11}.)


\end{proof}


\section{Sanity Check-3 : 4D free $U(1)$ theory under the Lorentz gauge}
\label{u(1) example}


\section{Rigorous Derivation of Chiral Anomaly}

In this section, we rigorously justify Fujikawa's computation of chiral anomaly.



\section{Relation to non-local formalisms}

In~\cite[Ch.8]{Sei82}, it is pointed out that non-local objects, such as Wilson loops, provide a more natural framework for gauge theories as they are better suited for describing behaviors of particles in such theories. 

In fact, the list of Euclidean axioms for Schwinger functions defined in terms of Wilson loops is presented in~\cite[p.164]{Sei82} and~\cite{Thie97}. We aim to show that such ``non-local''  Schwinger functions can be constructed from ``local'' ones satisfying the axioms stated in Sec.~\ref{local axioms}.

\section{Conclusion}

The next natural step is to proceed toward Wightman functions on the Minkowski spacetime via analytic continuation in analogy to \cite{OstSch73, OstSch75}. The Wightman Reconstruction Theorem~\cite[p,117 Theorem 3.7]{StrWig64} or reconstruction of time-ordered products~\cite{time79} will be modified for gauge theories as well. Lastly, we aim to establish connection between our extended Wightman axioms and the existing AQFT formalism for gauge theories.

We are deeply grateful for Professors Arthur Jaffe, Jürg Fröhlich, Erhard Seiler, Klaus Fredenhagen, Martin Hairer, Abdelmalek Abdesselam, and Iosif Pinelis for their valuable helps and insights.

\appendix
\section{Miscellaneous Calculations for Section \ref{motive heuristics}.}
\label{appendix:a}

In this appendix, we present detailed justifications for the claims made in Section \ref{motive heuristics}.

\addtocontents{toc}{\protect\setcounter{tocdepth}{1}}

\subsection*{Proof of Consistency Check \ref{consistency check1}}

\begin{itemize}
    \item For $\bigl(g \cdot S- g \cdot C\bigr)_{(\alpha,i)}$ in item (\ref{1 claim}),

\begin{subequations}
    \label{sch colocating behavior check0}
    \begin{align*}
       &\lim\limits_{n \to \infty}  \int_{(\mathbb{R}^4)^2} t_{\alpha}\bigl(g \cdot S- g \cdot C\bigr)_{(\alpha,i)}(\undertilde{x}) f(\undertilde{y}) \bigl[A\Delta_n(h) \bigr](\undertilde{x},\undertilde{y}) dxdy \tag{\ref{sch colocating behavior check0}} \\
       &= \lim\limits_{n \to \infty}   \int_{(\mathbb{R}^4)^2}  \bigl(S-C \bigr)_{(\alpha,i)}(\undertilde{x})g(\undertilde{x})t_{\alpha}g^{-1}(\undertilde{x}) f(\undertilde{y}) \bigl[A\Delta_n(h) \bigr](\undertilde{x},\undertilde{y})dxdy \tag{\ref{sch colocating behavior check0}a}\\
       &= \int_{\mathbb{R}^4} \bigl(S-C \bigr)_{(\alpha,i)}(\undertilde{x})g(\undertilde{x})t_{\alpha}g^{-1}(\undertilde{x})A(\undertilde{x},\undertilde{x})f(\undertilde{x})h(\undertilde{x})dx  \tag{\ref{sch colocating behavior check0}b}\\
       &=  t_{\alpha}\bigl(g \cdot S- g \cdot C \bigr)_{(\alpha,i)}\bigl( A \bigl \lvert_{\text{diag}} f h  \bigr) \tag{\ref{sch colocating behavior check0}c}
    \end{align*}
\end{subequations}
where passing from (\ref{sch colocating behavior check0}a) to (\ref{sch colocating behavior check0}b) is justified by the definitions of $A$ and $g$ combined with~\eq{sch colocating behavior0}. 
\end{itemize}

\begin{itemize}
     \item For $\bigl(g \cdot S- g \cdot C \bigr)_{(\alpha,i),(\beta,j)}$ in item (\ref{1 claim}),
\begin{subequations}
 \label{sch colocating behavior check}
\begin{align*}
       &\lim\limits_{n \to \infty}  \int_{(\mathbb{R}^4)^3} t_{\alpha}\otimes t_{\beta}\bigl(g \cdot S- g \cdot C \bigr)_{(\alpha,i),(\beta,j)}(\undertilde{x}, \undertilde{y})  f(\undertilde{z})w(\undertilde{x}) \bigl[A\Delta_n(h) \bigr](\undertilde{y},\undertilde{z}) \tag{\ref{sch colocating behavior check}} \\
     &= \lim\limits_{n \to \infty} \int_{(\mathbb{R}^4)^3}\bigl( S-  C \bigr)_{(\alpha,i),(\beta,j)}(\undertilde{x}, \undertilde{y}) f(\undertilde{z}) w(\undertilde{x})g(\undertilde{x})t_{\alpha}g^{-1}(\undertilde{x}) \otimes g(\undertilde{y})t_{\beta}g^{-1}(\undertilde{y})\bigl[A\Delta_n(h) \bigr](\undertilde{y},\undertilde{z}) \tag{\ref{sch colocating behavior check}a}\\
     &- \lim\limits_{n \to \infty}  \Biggl( \bigl( S-  C \bigr)_{(\alpha,i)}( wgt_{\alpha}g^{-1}) \otimes \int_{(\mathbb{R}^4)^2}\bigl[(\partial_j g)g^{-1} \bigr](\undertilde{y}) f(\undertilde{z})\bigl[A\Delta_n(h) \bigr](\undertilde{y},\undertilde{z})dydz \Biggr) \tag{\ref{sch colocating behavior check}b}\\
     &- \lim\limits_{n \to \infty}  \Biggl( \int_{\mathbb{R}^4} w(\undertilde{x})\bigl[(\partial_i g)g^{-1} \bigr](\undertilde{x}) dx \otimes \int_{(\mathbb{R}^4)^2}\bigl( S-  C \bigr)_{(\beta,j)}(\undertilde{y}) f(\undertilde{z})g(\undertilde{y})t_{\beta}g^{-1}(\undertilde{y})\bigl[A\Delta_n(h) \bigr](\undertilde{y},\undertilde{z})dydz\Biggr) \tag{\ref{sch colocating behavior check}c}\\
     &=\int_{(\mathbb{R}^4)^2}\bigl( S-  C \bigr)_{(\alpha,i),(\beta,j)}(\undertilde{x},\undertilde{y})(w gt_\alpha g^{-1})(\undertilde{x}) \otimes (fhgt_\beta g^{-1})(\undertilde{y})A(\undertilde{y},\undertilde{y})dxdy \tag{\ref{sch colocating behavior check}d}\\
     &-  \bigl( S-  C \bigr)_{(\alpha,i)}( wgt_{\alpha}g^{-1}) \otimes \int_{\mathbb{R}^4}\bigl[(\partial_j g)g^{-1} \bigr](\undertilde{y}) f(\undertilde{y}) h(\undertilde{y})A(\undertilde{y},\undertilde{y})dy \tag{\ref{sch colocating behavior check}e}\\
     &-\int_{\mathbb{R}^4} w(\undertilde{x})\bigl[(\partial_i g)g^{-1} \bigr](\undertilde{x}) dx \otimes \bigl( S-  C \bigr)_{(\beta,j)}\bigl(A \bigl \lvert_{\text{diag}} fhgt_\beta g^{-1} \bigr) \tag{\ref{sch colocating behavior check}f}\\
     &=  t_\alpha \otimes t_\beta \bigl( g \cdot S-g \cdot C\bigr)_{(\alpha,i), (\beta, j)}\Bigl( w \otimes A \bigl \lvert_{\text{diag}} fh \Bigr)   \tag{\ref{sch colocating behavior check}g}
  \end{align*}
    \end{subequations}
where~\eq{sch colocating behavior} is used to go from (\ref{sch colocating behavior check}a) to (\ref{sch colocating behavior check}d) and~\eq{sch colocating behavior0} to go from (\ref{sch colocating behavior check}c) to (\ref{sch colocating behavior check}f). 

Also,
\begin{subequations}
 \label{sch colocating behavior check11}
\begin{align*}
       &\lim\limits_{n,n' \to \infty}  \int_{(\mathbb{R}^4)^4} t_{\alpha}\otimes t_{\beta}\bigl(g \cdot S- g \cdot C \bigr)_{(\alpha,i),(\beta,j)}(\undertilde{x}, \undertilde{y}) f(\undertilde{x'})w(\undertilde{y'}) \bigl[A\Delta_n(h) \bigr](\undertilde{x},\undertilde{x'}) \bigl[A'\Delta'_{n'}(s) \bigr](\undertilde{y},\undertilde{y'}) \tag{\ref{sch colocating behavior check11}} \\
     &= \lim\limits_{n,n' \to \infty}  \int_{(\mathbb{R}^4)^4} \bigl( S-  C \bigr)_{(\alpha,i),(\beta,j)}(\undertilde{x}, \undertilde{y}) f(\undertilde{x'})w(\undertilde{y'}) \bigl[A\Delta_n(h) \bigr](\undertilde{x},\undertilde{x'})(gt_{\alpha} g^{-1})(\undertilde{x})\otimes\\
     &  \bigl[A'\Delta'_{n'}(s) \bigr](\undertilde{y},\undertilde{y'}) (gt_{\beta} g^{-1})(\undertilde{y}) \tag{\ref{sch colocating behavior check11}a}\\\\
     &-\lim\limits_{n \to \infty} \Biggl( \int_{(\mathbb{R}^4)^2} \bigl(S-C \bigr)_{(\alpha,i)}(\undertilde{x})f(\undertilde{x'})  \bigl[A\Delta_n(h) \bigr](\undertilde{x},\undertilde{x'}) (gt_{\alpha} g^{-1})(\undertilde{x})\Biggr) \otimes \\
&\lim\limits_{n' \to \infty}\Biggl(\int_{(\mathbb{R}^4)^2} w(\undertilde{y'})\bigl[A'\Delta'_{n'}(s) \bigr](\undertilde{y},\undertilde{y'}) \bigl[ (\partial_j g)g^{-1}](\undertilde{y})\Biggr) \tag{\ref{sch colocating behavior check11}b}\\\\
&- \lim\limits_{n \to \infty}\Biggl(\int_{(\mathbb{R}^4)^2} f(\undertilde{x'})\bigl[A\Delta_{n}(h) \bigr](\undertilde{x},\undertilde{x'}) \bigl[ (\partial_i g)g^{-1}](\undertilde{x})\Biggr) \otimes \\
&\lim\limits_{n' \to \infty}\Biggl( \int_{(\mathbb{R}^4)^2} \bigl(S-C \bigr)_{(\beta,j)}(\undertilde{y})w(\undertilde{y'})  \bigl[A'\Delta'_{n'}(s) \bigr](\undertilde{y},\undertilde{y'}) (gt_{\beta} g^{-1})(\undertilde{y})\Biggr)\tag{\ref{sch colocating behavior check11}c}\\\\
&=\bigl(S-C\bigr)_{(\alpha,i),(\beta,j)}\Bigl( A \bigl \lvert_{\text{diag}} fh gt_\alpha g^{-1} \otimes A' \bigl \lvert_{\text{diag}}   ws gt_\beta g^{-1} \Bigr) \tag{\ref{sch colocating behavior check11}d} \\
&- \bigl(S-C\bigr)_{(\alpha,i)}\Bigl( A \bigl \lvert_{\text{diag}} fh gt_\alpha g^{-1}\Bigr) \otimes \Biggl( \int_{\mathbb{R}^4}w(\undertilde{y})A'(\undertilde{y},\undertilde{y})s(\undertilde{y})\bigl[ (\partial_j g)g^{-1} \bigr](\undertilde{y}) \Biggr) \tag{\ref{sch colocating behavior check11}e}\\
&- \Biggl( \int_{\mathbb{R}^4}f(\undertilde{x})A(\undertilde{x},\undertilde{x})h(\undertilde{x})\bigl[ (\partial_i g)g^{-1} \bigr](\undertilde{x}) \Biggr) \otimes \bigl(S-C\bigr)_{(\beta,j)}\Bigl( A' \bigl \lvert_{\text{diag}} ws gt_\beta g^{-1}\Bigr) \tag{\ref{sch colocating behavior check11}f}\\\\
&=  t_\alpha \otimes t_\beta \bigl( g \cdot S- g \cdot C\bigr)_{(\alpha,i), (\beta, j)}\Bigl( A \bigl \lvert_{\text{diag}} fh \otimes  A' \bigl \lvert_{\text{diag}}ws \Bigr)  \tag{\ref{sch colocating behavior check11}g}
\end{align*}
\end{subequations}
where~\eq{sch colocating behavior11} is used to go from (\ref{sch colocating behavior check11}a) to (\ref{sch colocating behavior check11}d), while~\eq{sch colocating behavior0} is used to go from (\ref{sch colocating behavior check11}b) to (\ref{sch colocating behavior check11}e) and (\ref{sch colocating behavior check11}c) to (\ref{sch colocating behavior check11}f) respectively.
 \end{itemize}

 \begin{itemize}
   \item For $\bigl(g \cdot S- g \cdot C \bigr)_{(\alpha,i),(\beta,j),(\gamma,k)}$ in item (\ref{1 claim}), 
\begin{subequations}
\label{sch colocating behavior check1}
\begin{align}
 & \int_{(\mathbb{R}^4)^4} t_{\alpha} \otimes t_{\beta} \otimes t_{\gamma} f(\undertilde{x'})\bigl( g\cdot S -g \cdot C\bigr)_{(\alpha,i), (\beta, j),(\gamma,k)}(\undertilde{x}, \undertilde{y},\undertilde{z})  \bigl[A\Delta_n(h) \bigr](\undertilde{x},\undertilde{x'}) F(\undertilde{y},\undertilde{z}) \tag{\ref{sch colocating behavior check1}} \\
     & = \int_{(\mathbb{R}^4)^4} f(\undertilde{x'}) \bigl(  S -C\bigr)_{(\alpha,i), (\beta, j),(\gamma,k)}(\undertilde{x}, \undertilde{y},\undertilde{z}) (gt_\alpha g^{-1})(\undertilde{x}) \otimes (g t_\beta g^{-1})(\undertilde{y}) \otimes (gt_\gamma g^{-1})(\undertilde{z}) \bigl[A\Delta_n(h) \bigr](\undertilde{x},\undertilde{x'}) F(\undertilde{y},\undertilde{z})\\
     &- \int_{(\mathbb{R}^4)^4} f(\undertilde{x'}) \bigl(  S -C\bigr)_{(\beta, j),(\gamma,k)}( \undertilde{y},\undertilde{z}) \bigl[(\partial_i g)g^{-1} \bigr](\undertilde{x}) \otimes (gt_\beta g^{-1})(\undertilde{y}) \otimes (gt_\gamma g^{-1})(\undertilde{z}) \bigl[A\Delta_n(h) \bigr](\undertilde{x},\undertilde{x'}) F(\undertilde{y},\undertilde{z})\\
     &-\int_{(\mathbb{R}^4)^4} f(\undertilde{x'}) \bigl(  S -C\bigr)_{(\alpha, i),(\gamma,k)}( \undertilde{x},\undertilde{z}) (gt_\alpha g^{-1})(\undertilde{x})\otimes \bigl[(\partial_j g)g^{-1} \bigr](\undertilde{y}) \otimes (gt_\gamma g^{-1})(\undertilde{z}) \bigl[A\Delta_n(h) \bigr](\undertilde{x},\undertilde{x'}) F(\undertilde{y},\undertilde{z})\\
     &- \int_{(\mathbb{R}^4)^4} f(\undertilde{x'}) \bigl(  S -C\bigr)_{(\alpha, i),(\beta,j)}( \undertilde{x},\undertilde{y}) (gt_\alpha g^{-1})(\undertilde{x})  \otimes (gt_\beta g^{-1}) (\undertilde{y}) \otimes \bigl[(\partial_k g)g^{-1} \bigr](\undertilde{z})\bigl[A\Delta_n(h) \bigr](\undertilde{x},\undertilde{x'}) F(\undertilde{y},\undertilde{z})\\
     &+ \int_{(\mathbb{R}^4)^4} f(\undertilde{x'}) \bigl(  S -C\bigr)_{(\alpha, i)}( \undertilde{x}) (gt_\alpha g^{-1})(\undertilde{x})  \otimes\bigl[(\partial_j g)g^{-1} \bigr](\undertilde{y}) \otimes \bigl[(\partial_k g)g^{-1} \bigr](\undertilde{z})\bigl[A\Delta_n(h) \bigr](\undertilde{x},\undertilde{x'}) F(\undertilde{y},\undertilde{z})\\
     &+ \int_{(\mathbb{R}^4)^4} f(\undertilde{x'}) \bigl(  S -C\bigr)_{(\beta, j)}( \undertilde{y}) \bigl[(\partial_i g)g^{-1} \bigr](\undertilde{x}) \otimes (gt_\beta g^{-1})(\undertilde{y}) \otimes \bigl[(\partial_k g)g^{-1} \bigr](\undertilde{z}) \bigl[A\Delta_n(h) \bigr](\undertilde{x},\undertilde{x'}) F(\undertilde{y},\undertilde{z})\\
     &+ \int_{(\mathbb{R}^4)^4} f(\undertilde{x'}) \bigl(  S -C\bigr)_{(\gamma, k)}( \undertilde{z}) \bigl[(\partial_i g)g^{-1} \bigr](\undertilde{x}) \otimes \bigl[(\partial_j g)g^{-1} \bigr](\undertilde{y}) \otimes (gt_\gamma g^{-1})(\undertilde{z})\bigl[A\Delta_n(h) \bigr](\undertilde{x},\undertilde{x'}) F(\undertilde{y},\undertilde{z})\\
          \end{align}
  \end{subequations}
Thus, by taking the limit $n \to \infty$ of~\eq{sch colocating behavior check1}, we obtain
\begin{subequations}
\label{sch colocating behavior check11}
\begin{align}
& \lim\limits_{n \to \infty} \int_{(\mathbb{R}^4)^4} t_{\alpha} \otimes t_{\beta} \otimes t_{\gamma} f(\undertilde{x'})\bigl( g\cdot S -g \cdot C\bigr)_{(\alpha,i), (\beta, j),(\gamma,k)}(\undertilde{x}, \undertilde{y},\undertilde{z})  \bigl[A\Delta_n(h) \bigr](\undertilde{x},\undertilde{x'}) F(\undertilde{y},\undertilde{z}) \tag{\ref{sch colocating behavior check11}}\\
&=\int_{(\mathbb{R}^4)^3} \bigl(S-C\bigr)_{(\alpha,i), (\beta, j),(\gamma,k)}(\undertilde{x}, \undertilde{y},\undertilde{z}) A(\undertilde{x},\undertilde{x})(fh g t_\alpha g^{-1})(\undertilde{x}) \otimes (gt_\beta g^{-1})(\undertilde{y}) \otimes (gt_\gamma g^{-1})(\undertilde{z}) F(\undertilde{y},\undertilde{z})\\
     &-\Biggl(\int_{\mathbb{R}^4}  A \bigl \lvert_{\text{diag}} fh(\partial_i g)g^{-1} \Biggr) \otimes \Bigg( \int_{(\mathbb{R}^4)^2}\bigl(S-C\bigr)_{(\beta,j),(\gamma,k)}(\undertilde{y},\undertilde{z})F(\undertilde{y},\undertilde{z})(gt_\beta g^{-1})(\undertilde{y}) \otimes (gt_\gamma g^{-1})(\undertilde{z}) \Biggr)\\
     &-\int_{(\mathbb{R}^4)^3}\bigl(S-C\bigr)_{(\alpha,i),(\gamma,k)}(\undertilde{x},\undertilde{z})(fhgt_\alpha g^{-1})(\undertilde{x})A(\undertilde{x},\undertilde{x}) \otimes \bigl[ (\partial_j g)g^{-1}\bigr](\undertilde{y}) \otimes (gt_\gamma g^{-1})(\undertilde{z}) F(\undertilde{y},\undertilde{z})\\
     &-\int_{(\mathbb{R}^4)^3}\bigl(S-C\bigr)_{(\alpha,i),(\beta,j)}(\undertilde{x},\undertilde{y}) (fhgt_\alpha g^{-1})(\undertilde{x})A(\undertilde{x},\undertilde{x}) \otimes (gt_\beta g^{-1})(\undertilde{y}) \otimes \bigl[ (\partial_k g)g^{-1}\bigr](\undertilde{z})F(\undertilde{y},\undertilde{z})\\
     &+\bigl(S-C\bigr)_{(\alpha,i)}\bigl( A \bigl \lvert_{\text{diag}} fh gt_\alpha g^{-1} \bigr) \otimes \Biggl( \int_{(\mathbb{R}^4)^2} \bigl[ (\partial_j g) g^{-1}\bigr](\undertilde{y}) \otimes \bigl[ (\partial_k g) g^{-1}\bigr](\undertilde{z})F(\undertilde{y},\undertilde{z}) \Biggr)\\
     &+\Biggl(\int_{\mathbb{R}^4} A \bigl \lvert_{\text{diag}} f h (\partial_i g) g^{-1} \Biggr) \otimes \Biggl( \int_{(\mathbb{R}^4)^2} \bigl( S- C\bigr)_{(\beta,j)}(\undertilde{y}) (gt_\beta g^{-1})(\undertilde{y}) \otimes \bigl[ (\partial_k g) g^{-1}\bigr](\undertilde{z}) F(\undertilde{y},\undertilde{z}) \Biggr)\\
     &+ \Biggl(\int_{\mathbb{R}^4} A \bigl \lvert_{\text{diag}} f h (\partial_i g) g^{-1} \Biggr) \otimes \Biggl( \int_{(\mathbb{R}^4)^2} \bigl( S- C\bigr)_{(\gamma,k)}(\undertilde{z}) \bigl[ (\partial_j g) g^{-1}\bigr](\undertilde{y}) \otimes (gt_\gamma g^{-1})(\undertilde{z})   F(\undertilde{y},\undertilde{z}) \Biggr)\\
     &=\int_{(\mathbb{R}^4)^3} t_\alpha \otimes t_\beta \otimes t_\gamma \bigl( g \cdot S- g \cdot C \bigr)_{(\alpha,i),(\beta,j),(\gamma,k)}(\undertilde{x},\undertilde{y},\undertilde{z})f(\undertilde{x})h(\undertilde{x})A(\undertilde{x},\undertilde{x})F(\undertilde{y},\undertilde{z})dxdydz
   \end{align}
  \end{subequations}
where~\eq{sch colocating behavior0}, (\ref{sch colocating behavior}) and (\ref{sch colocating behavior1}) are all used here to compute the limits.
 \end{itemize}

\begin{itemize}
    
\item To verify~\eq{FF limit consistency 0} in item (\ref{2 claim}), we compute as follows:
\begin{equation}
    \begin{split}
        \label{consistency 0 compute}
&t_{\alpha}\otimes t_{\beta}\Biggl(  \bigl[ g \cdot  S \bigr]_{(\alpha,i), (\beta,j)}- \bigl[ g \cdot  C \bigr]_{ (\alpha,i), (\beta,j), }\Biggr)\bigl(    A\Delta_{n}(h)\bigr)   \\
& =\Bigl(S_{(\alpha,i), (\beta,j)}-C_{(\alpha,i), (\beta,j)}\Bigr)\bigl( A gt_{\alpha}g^{-1}\otimes gt_{\beta}g^{-1}\Delta_{n}(h) \bigr)\\
&-\int_{(\mathbb{R}^4)^2} \bigl(S- C\bigr)_{(\alpha,i)}(\undertilde{x})g(\undertilde{x})t_{\alpha}g^{-1}(\undertilde{x}) \otimes \bigl[(\partial_j g)g^{-1}\bigr](\undertilde{y})\bigl[A\Delta_n(h)\bigr](\undertilde{x},\undertilde{y})dxdy\\
&-\int_{(\mathbb{R}^4)^2} \bigl(S- C\bigr)_{(\beta,j)}(\undertilde{y}) \bigl[A\Delta_n(h)\bigr](\undertilde{x},\undertilde{y})\bigl[(\partial_i g)g^{-1}\bigr](\undertilde{x}) \otimes g(\undertilde{y})t_{\beta}g^{-1}(\undertilde{y})  dxdy
    \end{split}
\end{equation}
When taking $n \to \infty$, the second line of~\eq{consistency 0 compute} has a limit in the desired way due to the assumption on $S_{(\alpha,i), (\beta,j)}-C_{(\alpha,i), (\beta,j)}$ as in~\eq{FF limit 0}. The third line of~\eq{consistency 0 compute} satisfies
\begin{equation*}
    \begin{split}
    &\lim\limits_{n \to \infty} \int_{(\mathbb{R}^4)^2} \bigl(S- C\bigr)_{(\alpha,i)}(\undertilde{x})g(\undertilde{x})t_{\alpha}g^{-1}(\undertilde{x}) \otimes \bigl[(\partial_j g)g^{-1}\bigr](\undertilde{y})\bigl[A\Delta_n(h)\bigr](\undertilde{x},\undertilde{y})dxdy \\
    &= \bigl(S- C\bigr)_{(\alpha,i)}\Bigl(  \bigl[A  gt_{\alpha}g^{-1} \otimes (\partial_j g)g^{-1}\bigr]   \Bigl \lvert_{\text{diag}}  h\Bigr)
    \end{split}
\end{equation*}
where 
\[
\bigl[A  gt_{\alpha}g^{-1} \otimes (\partial_j g)g^{-1}\bigr] \Bigl \lvert_{\text{diag}}(\undertilde{x})=A(\undertilde{x},\undertilde{x}) g(\undertilde{x})t_{\alpha}g^{-1}(\undertilde{x}) \otimes \bigl[(\partial_j g)g^{-1}](\undertilde{x})
\]
according to~\eq{sch colocating behavior0}. A similar limit can be obtained for the fourth line of~\eq{consistency 0 compute}:
\begin{equation*}
    \begin{split}
    &\lim\limits_{n \to \infty} \int_{(\mathbb{R}^4)^2} \bigl(S- C\bigr)_{(\beta,j)}(\undertilde{y}) \bigl[A\Delta_n(h)\bigr](\undertilde{x},\undertilde{y})\bigl[(\partial_i g)g^{-1}\bigr](\undertilde{x}) \otimes g(\undertilde{y})t_{\beta}g^{-1}(\undertilde{y})  dxdy \\
    &= \bigl(S- C\bigr)_{(\beta,j)}\Bigl(  \bigl[A   (\partial_i g)g^{-1} \otimes gt_{\beta}g^{-1}\bigr]   \Bigl \lvert_{\text{diag}}  h\Bigr)
    \end{split}
\end{equation*}
where
\[
\bigl[A   (\partial_i g)g^{-1} \otimes gt_{\beta}g^{-1}\bigr]   \Bigl \lvert_{\text{diag}}(\undertilde{x})=A(\undertilde{x},\undertilde{x}) \bigl[(\partial_i g)g^{-1}](\undertilde{x}) 
\otimes g(\undertilde{x})t_{\beta}g^{-1}(\undertilde{x}).
\]
\end{itemize}

\begin{itemize}

\item To verify~\eq{FF limit consistency 1} in item (\ref{2 claim}), we expand as follows:
\begin{subequations}
\label{FF consistency 1 check}
\begin{align}
   &t_\alpha \otimes t_{\beta'} \otimes t_{\gamma'} \bigl(   g \cdot  S  -g \cdot C \bigr)_{(\alpha,j), (\beta',i'), (\gamma',j')}\bigl( f \otimes  A'\Delta'_{n'}(h)\bigr) \tag{\ref{FF consistency 1 check}}\\
   &=\bigl( S-C\bigr)_{(\alpha,j), (\beta',i'), (\gamma',j')}\bigl( fgt_\alpha g^{-1} \otimes A' \Delta'_{n'}(h) \bigl( gt_{\beta'} g^{-1} \otimes gt_{\gamma'} g^{-1}\bigr) \bigr)  \label{A} \\
   &-\int_{(\mathbb{R}^4)^3} \bigl( S-C\bigr)_{(\alpha,j),(\beta',i')}(\undertilde{x},\undertilde{x'}) (f gt_\alpha g^{-1})(\undertilde{x}) \otimes \bigl[A'\Delta'_{n'}(h) \bigr](\undertilde{x'},\undertilde{y'}) (gt_{\beta'}g^{-1})(\undertilde{x'}) \otimes  \bigl[(\partial_{j'}g)g^{-1} \bigr](\undertilde{y'})  \label{B} \\
   &-\int_{(\mathbb{R}^4)^3} \bigl( S-C\bigr)_{(\alpha,j), (\gamma',j')}(\undertilde{x},\undertilde{y'})(f gt_\alpha g^{-1})(\undertilde{x})\otimes  \bigl[A'\Delta'_{n'}(h)\bigr](\undertilde{x'},\undertilde{y'})\bigl[(\partial_{i'}g)g^{-1}\bigr](\undertilde{x'})\otimes  ( gt_{\gamma'} g^{-1})(\undertilde{y'}) \label{C}\\
   &- \Biggl(\int_{\mathbb{R}^4} f (\partial_j g)g^{-1} \Biggr)\otimes \Biggl( \bigl(S-C\bigr)_{(\beta',i'), (\gamma', j')}\bigl(A' \Delta_{n'}(h) g t_{\beta'} g^{-1} \otimes gt_{\gamma'}g^{-1}\bigr) \Biggl) \label{D}\\
   & + \bigl(S-C\bigr)_{(\alpha,j)}\bigl(fg t_\alpha g^{-1} \bigr) \otimes \Biggl( \int_{(\mathbb{R}^4)^2} \bigl[A'\Delta'_{n'}(h) \bigr](\undertilde{x'},\undertilde{y'})\bigl[(\partial_{i'}g)g^{-1} \bigr](\undertilde{x'})\otimes \bigl[(\partial_{j'} g)g^{-1}  \bigr](\undertilde{y'})\Biggr)  \label{E}\\
   &+\Biggl( \int_{\mathbb{R}^4}  f \bigl( \partial_j g)g^{-1} \Biggr) \otimes \Biggl( \int_{(\mathbb{R}^4)^2} \bigl(S-C\bigr)_{(\beta',i')}(\undertilde{x'})\bigl[A' \Delta'_{n'}(h) \bigr](\undertilde{x'},\undertilde{y'})(g t_{\beta'}g^{-1})(\undertilde{x'})\otimes \bigl[(\partial_{j'}g) g^{-1}\bigr](\undertilde{y'}) \Biggr)  \label{F}\\
   &+\Biggl( \int_{\mathbb{R}^4}  f \bigl( \partial_j g)g^{-1} \Biggr) \otimes \Biggl(  \int_{(\mathbb{R}^4)^2}\bigl(S-C\bigr)_{(\gamma',j')}(\undertilde{y'})\bigl[A' \Delta'_{n'}(h) \bigr](\undertilde{x'},\undertilde{y'})\bigl[(\partial_{i'}g) g^{-1} \bigr](\undertilde{x'})\otimes (g t_{\beta'}g^{-1} )(\undertilde{y'})\Biggr) \label{G}
\end{align}
\end{subequations}
The limits of (\ref{A}) and (\ref{D}) as $n' \to \infty$ exist in the desired way due to the assumption~\eq{FF limit 1} and~\eq{FF limit 0} respectively. The limits of (\ref{B}) and (\ref{C}) are addressed by~\eq{sch colocating behavior}, while the limits of (\ref{F}) and (\ref{G}) are addressed by~\eq{sch colocating behavior0}. Lastly, the limit of (\ref{E}) is trivial. 
\end{itemize}

\begin{itemize}
    \item To show compatibility of~\eq{FF limit consistency 1} and~\eq{sch colocating behavior check1} in the form of~\eq{odd number pair}, we compute 
\begin{subequations}
\label{compatibility check odd}
\begin{align*}
         &\lim\limits_{n,n' \to \infty}  \int_{(\mathbb{R}^4)^4} f(\undertilde{x})\bigl(  g \cdot S- g \cdot C\bigr)_{(\alpha,j), (\beta',i'), (\gamma',j')}(\undertilde{y}, \undertilde{x'},\undertilde{y'}) \bigl[A\Delta_n(h) \bigr](\undertilde{x},\undertilde{y})\bigl[A'\Delta'_{n'}(w) \bigr](\undertilde{x'},\undertilde{y'}) \tag{\ref{compatibility check odd}}  \\
         & = ddd \\
         &= ddd \\
     &=  \lim\limits_{n' \to \infty} \bigl(  g \cdot S- g \cdot C\bigr)_{(\alpha,j), (\beta',i'), (\gamma',j')}\bigl( A\bigl\lvert_{\text{diag}} fh \otimes  A'\Delta'_{n'}(w)\bigr).
    \end{align*}
\end{subequations}
\end{itemize}

\begin{itemize}
    \item Lastly, to verify~\eq{FF limit consistency 2} in item (\ref{2 claim}), let us expand
\begin{subequations}
\label{consistency 2 midcompute}
\begin{align*}
   & t_\beta \otimes t_{\gamma} \otimes t_{\beta'} \otimes t_{\gamma'}\bigl(  g \cdot  S -  g \cdot  C \bigr)_{(\beta,i), (\gamma, j), (\beta',i'), (\gamma',j')}( A \Delta_n(f) \otimes A'\Delta'_{n'}(h)) \tag{\ref{consistency 2 midcompute}}\\\\
   &=\bigl(S-C\bigr)_{(\beta,i), (\gamma, j), (\beta',i'), (\gamma',j')}\Bigl( \bigl[ A \Delta_n(f) g t_\beta g^{-1} \otimes gt_\gamma g^{-1} \bigr] \otimes \bigl[ A'\Delta'_{n'}(h) g t_{\beta'} g^{-1} \otimes gt_{\gamma'} g^{-1} \bigr] \Bigr) \tag{\ref{consistency 2 midcompute}a}\\\\
   &-\int_{(\mathbb{R}^4)^4} \bigl(S-C\bigr)_{ (\gamma, j), (\beta',i'), (\gamma',j')}(\undertilde{y},\undertilde{x'},\undertilde{y'})\bigl[ A \Delta_n(f) \bigr](\undertilde{x},\undertilde{y})\bigl[ (\partial_i g)g^{-1}](\undertilde{x}) \otimes (gt_\gamma g^{-1})(\undertilde{y}) \otimes \\
   &\bigl[ A'\Delta'_{n'}(h) g t_{\beta'} g^{-1} \otimes gt_{\gamma'} g^{-1} \bigr](\undertilde{x'},\undertilde{y'}) \tag{\ref{consistency 2 midcompute}b}\\\\
   &-\int_{(\mathbb{R}^4)^4}  \bigl(S-C\bigr)_{ (\beta, i), (\beta',i'), (\gamma',j')}(\undertilde{x},\undertilde{x'},\undertilde{y'})\bigl[ A \Delta_n(f) \bigr](\undertilde{x},\undertilde{y}) (gt_\beta g^{-1})(\undertilde{x}) \otimes \bigl[ (\partial_j g)g^{-1}](\undertilde{y}) \otimes \\
   &\bigl[ A'\Delta'_{n'}(h) g t_{\beta'} g^{-1} \otimes gt_{\gamma'} g^{-1} \bigr](\undertilde{x'},\undertilde{y'}) \tag{\ref{consistency 2 midcompute}c}\\\\
   &-\int_{(\mathbb{R}^4)^4} \bigl(S-C\bigr)_{ (\beta, i), (\gamma,j), (\gamma',j')}(\undertilde{x},\undertilde{y},\undertilde{y'}) \bigl[ A\Delta_{n}(f) g t_{\beta} g^{-1} \otimes gt_{\gamma} g^{-1} \bigr](\undertilde{x},\undertilde{y}) \otimes \bigl[ (\partial_{i'}g)g^{-1}](\undertilde{x'}) \otimes \\
   & (gt_{\gamma'}g^{-1})(\undertilde{y'})\bigl[A'\Delta'_{n'}(h) \bigr](\undertilde{x'},\undertilde{y'}) \tag{\ref{consistency 2 midcompute}d}\\\\
   &- \int_{(\mathbb{R}^4)^4} \bigl(S-C\bigr)_{ (\beta, i), (\gamma,j), (\beta',i')}(\undertilde{x},\undertilde{y},\undertilde{y'}) \bigl[ A\Delta_{n}(f) g t_{\beta} g^{-1} \otimes gt_{\gamma} g^{-1} \bigr](\undertilde{x},\undertilde{y}) \otimes(gt_{\beta'}g^{-1})(\undertilde{x'}) \otimes \\
   & \bigl[ (\partial_{j'}g)g^{-1} \bigr](\undertilde{y'})\bigl[A'\Delta'_{n'}(h) \bigr](\undertilde{x'},\undertilde{y'})\tag{\ref{consistency 2 midcompute}e}\\\\
   &+ \bigl(S-C\bigr)_{(\beta,i),(\gamma,j)}\Bigl( A\Delta_{n}(f) g t_{\beta} g^{-1} \otimes gt_{\gamma} g^{-1}\Bigr) \otimes \\
   &\Biggl(\int_{(\mathbb{R}^4)^2} \bigl[A'\Delta'_{n'}(h) \bigr](\undertilde{x'},\undertilde{y'})\bigl[ (\partial_{i'} g)g^{-1}\bigr](\undertilde{x'}) \otimes \bigl[ (\partial_{j'} g)g^{-1}\bigr](\undertilde{y'}) \Biggr)\tag{\ref{consistency 2 midcompute}f}\\\\
   &+ \int_{(\mathbb{R}^4)^4} \bigl(S-C\bigr)_{(\beta,i),(\beta',i')}(\undertilde{x},\undertilde{x'})\bigl[A\Delta_{n}(f)\bigr](\undertilde{x},\undertilde{y})(gt_{\beta}g^{1-})(\undertilde{x}) \otimes \bigl[(\partial_j g)g^{-1}](\undertilde{y}) \otimes\\
   & \bigl[A'\Delta'_{n'}(h) \bigr](\undertilde{x'},\undertilde{y'})(g t_{\beta'}g^{-1})(\undertilde{x'})\otimes \bigl[(\partial_{j'}g)g^{-1}](\undertilde{y'})\tag{\ref{consistency 2 midcompute}g}\\\\
   &+\int_{(\mathbb{R}^4)^4} \bigl(S-C\bigr)_{(\beta,i),(\gamma',j')}(\undertilde{x},\undertilde{y'})\bigl[A\Delta_{n}(f)\bigr](\undertilde{x},\undertilde{y})(gt_\beta g^{-1})(\undertilde{x}) \otimes \bigl[ (\partial_j g)g^{-1}](\undertilde{y}) \otimes \\
   & \bigl[A'\Delta'_{n'}(h) \bigr](\undertilde{x'},\undertilde{y'}) \bigl[ (\partial_{i'}g)g^{-1}](\undertilde{x'}) \otimes (gt_{\gamma'}g^{-1})(\undertilde{y'})\tag{\ref{consistency 2 midcompute}h}\\\\
   &+\int_{(\mathbb{R}^4)^4} \bigl(S-C\bigr)_{(\gamma,j),(\beta',i')}(\undertilde{y},\undertilde{x'}) \bigl[A\Delta_{n}(f)\bigr](\undertilde{x},\undertilde{y}) \bigl[ (\partial_i g)g^{-1}\bigr](\undertilde{x}) \otimes (gt_\gamma g^{-1})(\undertilde{y}) \otimes \\
   & \bigl[A'\Delta'_{n'}(h) \bigr](\undertilde{x'},\undertilde{y'}) (gt_{\beta'}g^{-1})(\undertilde{x'}) \otimes \bigl[ (\partial_{j'} g)g^{-1}](\undertilde{y'})\tag{\ref{consistency 2 midcompute}i}\\\\
   &+\int_{(\mathbb{R}^4)^4} \bigl(S-C\bigr)_{(\gamma,j),(\gamma',j')}(\undertilde{y},\undertilde{y'}) \bigl[A\Delta_{n}(f)\bigr](\undertilde{x},\undertilde{y}) \bigl[ (\partial_i g)g^{-1}\bigr](\undertilde{x}) \otimes (gt_\gamma g^{-1})(\undertilde{y}) \otimes \\
   & \bigl[A'\Delta'_{n'}(h) \bigr](\undertilde{x'},\undertilde{y'}) \bigl[(\partial_{i'} g)g^{-1} \bigr](\undertilde{x'}) \otimes (gt_{\gamma'}g^{-1})(\undertilde{y'})\tag{\ref{consistency 2 midcompute}j}\\\\
   &+\Biggr(\int_{(\mathbb{R}^4)^2}  \bigl[A\Delta_{n}(f) \bigr](\undertilde{x},\undertilde{y})\bigl[ (\partial_i g)g^{-1}](\undertilde{x}) \otimes \bigl[ (\partial_j g)g^{-1}](\undertilde{y}) \Biggl) \otimes \\
   &\bigl(S-C\bigr)_{(\beta',i'),(\gamma',j')}\Bigl( A'\Delta'_{n'}(h) gt_{\beta'}g^{-1} \otimes gt_{\gamma'}g^{-1} \Bigr)\tag{\ref{consistency 2 midcompute}k}\\\\
   &-\Biggl( \int_{(\mathbb{R}^4)^2} \bigl(S-C\bigr)_{(\beta,i)}(\undertilde{x}) \bigl[A\Delta_{n}(f)\bigr](\undertilde{x},\undertilde{y}) (gt_\beta g^{-1})(\undertilde{x})\otimes \bigl[ (\partial_j g)g^{-1}\bigr](\undertilde{y}) \Biggr) \otimes \\
   & \Biggl( \int_{(\mathbb{R}^4)^2} \bigl[A'\Delta'_{n'}(h) \bigr](\undertilde{x'},\undertilde{y'})\bigl[ (\partial_{i'} g)g^{-1}\bigr](\undertilde{x'}) \otimes \bigl[ (\partial_{j'} g)g^{-1}\bigr](\undertilde{y'}) \Biggr) \tag{\ref{consistency 2 midcompute}l}\\\\
   &-\Biggl( \int_{(\mathbb{R}^4)^2} \bigl(S-C\bigr)_{(\gamma,j)}(\undertilde{y}) \bigl[A\Delta_{n}(f)\bigr](\undertilde{x},\undertilde{y}) \bigl[ (\partial_{i} g)g^{-1}\bigr](\undertilde{x}) \otimes (g t_{\gamma} g^{-1})(\undertilde{y}) \Biggr) \otimes \\
   &\Biggl( \int_{(\mathbb{R}^4)^2} \bigl[A'\Delta'_{n'}(h) \bigr](\undertilde{x'},\undertilde{y'})\bigl[ (\partial_{i'} g)g^{-1}\bigr](\undertilde{x'}) \otimes \bigl[ (\partial_{j'} g)g^{-1}\bigr](\undertilde{y'}) \Biggr) \tag{\ref{consistency 2 midcompute}m}\\\\
   &-\Biggl( \int_{(\mathbb{R}^4)^2}  \bigl[A\Delta_{n}(f)\bigr](\undertilde{x},\undertilde{y}) \bigl[ (\partial_{i} g)g^{-1}\bigr](\undertilde{x}) \otimes \bigl[ (\partial_{j} g)g^{-1}\bigr](\undertilde{y}) \Biggr) \otimes \\
   & \Biggl( \int_{(\mathbb{R}^4)^2} \bigl(S-C\bigr)_{(\beta',i')}(\undertilde{x'})\bigl[A'\Delta'_{n'}(h) \bigr](\undertilde{x'},\undertilde{y'}) (g t_{\beta'} g^{-1})(\undertilde{x'})\otimes \bigl[ (\partial_{j'} g)g^{-1}\bigr](\undertilde{y'}) \tag{\ref{consistency 2 midcompute}n}\Biggr) \\\\
   &-\Biggl( \int_{(\mathbb{R}^4)^2} \bigl[A\Delta_{n}(f)\bigr](\undertilde{x},\undertilde{y})\bigl[ (\partial_{i} g)g^{-1}\bigr](\undertilde{x}) \otimes \bigl[ (\partial_{j} g)g^{-1}\bigr](\undertilde{y}) \Biggr) \otimes \\
   & \Biggl( \int_{(\mathbb{R}^4)^2} \bigl(S-C\bigr)_{(\gamma',j')}(\undertilde{y'})\bigl[A'\Delta'_{n'}(h) \bigr](\undertilde{x'},\undertilde{y'}) \bigl[ (\partial_{i'} g)g^{-1}\bigr](\undertilde{x'}) \otimes (g t_{\gamma'} g^{-1})(\undertilde{y'})\Biggr) \tag{\ref{consistency 2 midcompute}o}\\
\end{align*}
\end{subequations}
Let us anaylze each term in~\eq{consistency 2 midcompute}.
\begin{itemize}[label=$\diamond$]
   
    \item The limit of (\ref{consistency 2 midcompute}a) as $n,n' \to \infty$ exists in the desired way due to~\eq{FF limit 2}.

    \item The limits of (\ref{consistency 2 midcompute}b) and (\ref{consistency 2 midcompute}c) as $n,n' \to \infty$ exist in the desired way due to~\eq{odd number pair}.

    \item The limits of (\ref{consistency 2 midcompute}d) and (\ref{consistency 2 midcompute}e) as $n,n' \to \infty$ exist in the desired way due to~\eq{odd number pair} and the assumption of symmetry under commutation of the indices and spacetime arguments. 

    \item The limits of (\ref{consistency 2 midcompute}f) and (\ref{consistency 2 midcompute}k) as $n,n' \to \infty$ exist in the desired way due to~\eq{FF limit 0}.

\item The limits of (\ref{consistency 2 midcompute}g), (\ref{consistency 2 midcompute}h), (\ref{consistency 2 midcompute}i) and (\ref{consistency 2 midcompute}j) as $n,n' \to \infty$ exist in the desired way due to~\eq{sch colocating behavior11}.

    \item The limits of (\ref{consistency 2 midcompute}l), (\ref{consistency 2 midcompute}m), (\ref{consistency 2 midcompute}n) and (\ref{consistency 2 midcompute}o) as $n,n' \to \infty$ exist in the desired way due to~\eq{sch colocating behavior0}.
\end{itemize}

\end{itemize}

\subsection*{Proof of dddd}

\section{More}
\label{appendix:b}

\bibliographystyle{unsrt}
\bibliography{refs}

\begin{thebibliography}{10}

\bibitem{OstSch73}
K.~Osterwalder and R.~Schrader.
\newblock Axioms for {E}uclidean {G}reen's functions.
\newblock {\em Commun. Math. Phys.}, 31:83--112, 1973.

\bibitem{OstSch75}
K.~Osterwalder and R.~Schrader.
\newblock Axioms for {E}uclidean {G}reen's functions. {I}{I}.
\newblock {\em Commun. Math. Phys.}, 42:281--305, 1975.

\bibitem{Nel73}
Edward Nelson.
\newblock Construction of {Q}uantum {F}ields from {M}arkoff {F}ields.
\newblock {\em Journal of Functional Analysis}, 12:97--112, 1973.

\bibitem{time79}
J.-P. Eckmann and H.~Epstein.
\newblock Time-{O}rdered {P}roducts and {S}chwinger {F}unctions.
\newblock {\em Commun. math. Phys.}, 64:95--130, 1979.

\bibitem{BryFroSei79}
D.~Brydges, J.~Frohlich, and E.~Seiler.
\newblock On the construction of quantized gauge fields, i. general results.
\newblock {\em Ann. Physics}, 121:227--284, 1979.

\bibitem{BryFroSei80}
D.~Brydges, J.~Frohlich, and E.~Seiler.
\newblock On the construction of quantized gauge fields, {I}{I}. convergence of the lattice approximation.
\newblock {\em Comm. Math. Phys.}, 71, 1980.

\bibitem{BryFroSei81}
D.~Brydges, J.~Frohlich, and E.~Seiler.
\newblock On the construction of quantized gauge fields, {I}{I}{I}. the two-dimensional abelian higgs model without cutoffs.
\newblock {\em Comm. Math. Phys.}, 79, 1981.

\bibitem{Dimoscalar}
J.~Dimock.
\newblock Nonperturbative renormalization of scalar {QED} in d=3.
\newblock {\em arXiv e-prints}, 2021.

\bibitem{Dim18}
J.~Dimock.
\newblock Ultraviolet stability for {QED} in d = 3.
\newblock {\em J. Math; Phys.}, 59, 2018.

\bibitem{Dim20}
J.~Dimock.
\newblock Multiscale block averaging in {QED} in d = 3.
\newblock {\em J. Math; Phys.}, 61, 2020.

\bibitem{Dim20a}
J.~Dimock.
\newblock Ultraviolet regularity for {QED} in d = 3.
\newblock {\em arXiv:1512.04373}, 2020.

\bibitem{Dimo1}
J.~Dimock.
\newblock {T}he renormalization group according to {B}alaban-{I}, small fields.
\newblock {\em Rev. Math. Phys.}, 25(1330010):1--64, 2013.

\bibitem{Dimo2}
J.~Dimock.
\newblock {T}he renormalization group according to {B}alaban-{II}, large fields.
\newblock {\em Rev. Math. Phys.}, 54(092301):1--85, 2013.

\bibitem{Dimo3}
J.~Dimock.
\newblock {T}he renormalization group according to {B}alaban-{III}, convergence.
\newblock {\em Annales Henri Poincare.}, 15:2133--2175, 2014.

\bibitem{Bal11}
T.~Balaban.
\newblock ({H}iggs)$_{2,3}$ quantum fields in a finite volume-{I}.
\newblock {\em Commun. Math. Phys.}, 85:603--636, 1982.

\bibitem{Bal12}
T.~Balaban.
\newblock ({Hi}ggs)$_{2,3}$ quantum fields in a finite volume-{II}.
\newblock {\em Commun. Math. Phys.}, 86:555--594, 1982.

\bibitem{Bal13}
T.~Balaban.
\newblock ({H}iggs)$_{2,3}$ quantum fields in a finite volume-{III}.
\newblock {\em Commun. Math. Phys.}, 88:411--445, 1983.

\bibitem{Bal14}
T.~Balaban.
\newblock Regularity and decay of lattice {G}reen's functions.
\newblock {\em Commun. Math. Phys.}, 89:571--597, 1983.

\bibitem{GroKinSen89}
Leonard Gross, Christopher King, and Ambar Sengupta.
\newblock {T}wo {D}imensional {Y}ang-{M}ills {T}heory via {S}tochastic {D}ifferential {E}quations.
\newblock {\em Annals of Physics}, 194:65--112, 1985.

\bibitem{stoc1}
Ajay Chandra, Ilya Chevyrev, Martin Hairer, and Hao Shen.
\newblock Langevin dynamic for the 2d yang-mills measure.
\newblock {\em Publications math{\'e}matiques de l'IH{\'E}S}, 136:1--147, 2022.

\bibitem{stoc2}
Martin~Hairer Ajay~Chandra, Ilya~Chevyrev and Hao Shen.
\newblock Stochastic quantisation of {Y}ang {M}ills {H}iggs in 3{D}.
\newblock {\em ArXiv e-prints}, 2022.

\bibitem{Dri89}
Bruce~K. Driver.
\newblock Ym$_2$ {C}ontinuum {E}xpectations, {L}attice {C}onvergence, and {L}assos.
\newblock {\em Commun. Math. Phys.}, 123:575--616, 1989.

\bibitem{jaffecommun}
Arthur Jaffe.
\newblock Private communication, 2024.

\bibitem{OstSei78}
K.~Osterwalder and E.~Seiler.
\newblock Gauge field theories on a lattice.
\newblock {\em ANNALS OF PHYSICS}, 110:440--471, 1978.

\bibitem{Jaf00}
Arthur Jaffe.
\newblock Constructive quantum field theory.
\newblock {\em Mathematical Physics}, pages 111--127, 2000.

\bibitem{Wei96}
S.~Weinberg.
\newblock {\em The Quantum Theory of Fields, Volume II}.
\newblock Cambridge University Press, Cambridge, 1996.

\bibitem{484755}
Liding~Yao (https://mathoverflow.net/users/118469/liding yao).
\newblock The space of bounded smooth functions with rapidly decaying derivatives.
\newblock MathOverflow.
\newblock URL:https://mathoverflow.net/q/484755 (version: 2025-02-20).

\bibitem{Treves67}
Francois Treves.
\newblock {\em Topological Vector Spaces, Distributions and Kernels}.
\newblock Academic Press Inc., New York, New York, 1967.

\bibitem{coinciding}
Roberto Fernandez, Jurg Frohlich, and Alan Sokal.
\newblock {\em Random Walks, Critical Phenomena, and Triviality in Quantum Field Theory}.
\newblock Springer-Verlag, New York, 1992.

\bibitem{Haa96}
Rudolf Haag.
\newblock {\em Local Quantum Physics : Fields, Particles, Algebras}.
\newblock Springer, New York, 1996.

\bibitem{BoLoOkTo90}
N.~N. Bogolubov, A.~A. Logubov, A.~I. Oksak, and I.~T. Todorov.
\newblock {\em General Principles of Quantum Field Theory}.
\newblock Kluwer Academic Publishers, Netherlands, 1990.

\bibitem{StrWig64}
R.~Streater and A.~Wightman.
\newblock {\em PCT, Spin and Statistics and all that}.
\newblock Princeton`University Press, Princeton, 1964.

\bibitem{Eli75}
S.~Elitzur.
\newblock Impossibility of spontaneously breaking local symmetries.
\newblock {\em Physical Review D}, Volume 12, Number 12, 1975.

\bibitem{FroMorStr81}
G.~Morchio J.~Frohlich and F.~Strocchi.
\newblock Higgs phenomenon without symmetry breaking order parameter.
\newblock {\em Nuclear Physics B}, 190[FS3]:553--582, 1981.

\bibitem{FadPop67}
L.~Faddeev and V.N. Popov.
\newblock Feynman diagrams for the {Y}ang-{M}ills field.
\newblock {\em Phys. Lett.}, 25B 29, 1967.

\bibitem{Fra08}
P.~Frampton.
\newblock {\em Gauge Field Theory}.
\newblock Wiley-VCH, Germany, 2008.

\bibitem{BruFreKoh96}
R.~Brunetti, K.~Fredenhagen, and M.~Köhler.
\newblock The {M}icrolocal {S}pectrum {C}ondition and {W}ick {P}olynomials of {F}ree {F}ields on {C}urved {S}pacetimes.
\newblock {\em Commun. Math. Phys}, 180, 1996.

\bibitem{Rud91}
W.~Rudin.
\newblock {\em Functional Analysis}.
\newblock McGraw-Hill, Singapore, 1991.

\bibitem{ReeSim72}
M.~Reed and B.~Simon.
\newblock {\em Functional Analysis, Vol. {I}}.
\newblock Academic Press, New York, 1972.

\bibitem{moment}
H.~J. Borchers and J.~Yngvason.
\newblock Necessary and sufficient conditions for integral representations of wightman functionals at schwinger points.
\newblock {\em Commun. math. Phys.}, 47:197--213, 1976.

\bibitem{moment2}
Jurg Frohlich.
\newblock Schwinger {F}unctions and {T}heir {G}enerating {F}unctionals. {I}{I}. {M}arkovian and {G}eneralized {P}ath {S}pace {M}easures on $\mathcal{S}'$.
\newblock {\em Advances in Mathematics}, 23:119--180, 1977.

\bibitem{MO11}
Abdelmalek~Abdesselam (https://mathoverflow.net/users/7410/abdelmalek abdesselam).
\newblock Moment problem, ergodicity and spectral gap on the space of tempered distributions.
\newblock MathOverflow.
\newblock URL:https://mathoverflow.net/q/478753 (version: 2024-09-13).

\bibitem{Sei82}
Erhard Seiler.
\newblock {\em Gauge Theories as a Problem of Constructive Quantum Field Theory and Statistical Mechanics}.
\newblock Springer, Berlin Heidelberg, 1982.

\bibitem{Thie97}
Thomas Thiemann.
\newblock An {A}xiomatic {A}pproach to {Q}uantum {G}auge {F}ield {T}heory.
\newblock {\em Symplectic Singularities and Geometry of Gauge Fields}, 39, 1997.

\end{thebibliography}

\end{document}